\documentclass[useAMS,usenatbib]{mn2e}
\usepackage{lineno}
\def\mnras{MNRAS}
\def\apj{ApJ}
\def\araa{ARA\&A}
\def\apjs{ApJS}

\usepackage{bbold}
\usepackage{bm}
\usepackage{graphicx}
\usepackage{float}
\usepackage{amssymb}
\usepackage{amsfonts}
\usepackage{amsmath}
\usepackage{color}
\usepackage{natbib}
\usepackage{hyperref}
\usepackage{longtable}
\usepackage{pdflscape}
\usepackage{dirtytalk}
\usepackage{multirow}
\usepackage{hhline}

\usepackage{mathtools}

\DeclarePairedDelimiter\abs{\lvert}{\rvert}%
\DeclarePairedDelimiter\norm{\lVert}{\rVert}%
\newcommand{\specialcell}[2][c]{%
  \begin{tabular}[#1]{@{}c@{}}#2\end{tabular}}

\newcommand*{\tran}{^{\mkern-1.5mu\mathsf{T}}}

\newcommand\pr{\mathbb P}
\DeclareMathOperator{\sigmoid}{sigmoid}

\usepackage{eso-pic}

\AddToShipoutPictureBG*{%
  \AtPageUpperLeft{%
    \hspace{0.75\paperwidth}%
    \raisebox{-3.5\baselineskip}{%
      \makebox[0pt][l]{\textnormal{DES-2015-0048}}
}}}%

\AddToShipoutPictureBG*{%
  \AtPageUpperLeft{%
    \hspace{0.75\paperwidth}%
    \raisebox{-4.5\baselineskip}{%
      \makebox[0pt][l]{\textnormal{FERMILAB-PUB-21-385-SCD}}
}}}%

\usepackage{xcolor}

\title[Machine finding quads]{Finding quadruply imaged quasars with machine learning. I. Methods}
\author[DES Collaboration]{
\parbox{\textwidth}{
\Large
A.~Akhazhanov,$^{1,2}$\thanks{e-mail:gkcalat@ucla.edu}
A.~More,$^{3,4}$
A.~Amini,$^{5}$
C.~Hazlett,$^{5,6}$
T.~Treu,$^{7}$\thanks{Packard fellow}
S.~Birrer,$^{8}$
A.~Shajib,$^{9}$\thanks{NHFP Einstein Fellow}
P.~Schechter,$^{10}$
C.~Lemon,$^{11}$
B.~Nord,$^{12,13,9}$
M.~Aguena,$^{14}$
S.~Allam,$^{12}$
F.~Andrade-Oliveira,$^{15,14}$
J.~Annis,$^{12}$
D.~Brooks,$^{16}$
E.~Buckley-Geer,$^{9,12}$
D.~L.~Burke,$^{17,18}$
A.~Carnero~Rosell,$^{14}$
M.~Carrasco~Kind,$^{19,20}$
J.~Carretero,$^{21}$
A.~Choi,$^{22}$
C.~Conselice,$^{23,24}$
M.~Costanzi,$^{25,26,27}$
L.~N.~da Costa,$^{14,28}$
M.~E.~S.~Pereira,$^{29,30}$
J.~De~Vicente,$^{31}$
S.~Desai,$^{32}$
J.~P.~Dietrich,$^{33}$
P.~Doel,$^{16}$
S.~Everett,$^{34}$
I.~Ferrero,$^{35}$
D.~A.~Finley,$^{12}$
B.~Flaugher,$^{12}$
J.~Frieman,$^{12,13}$
J.~Garc\'ia-Bellido,$^{36}$
D.~W.~Gerdes,$^{37,29}$
D.~Gruen,$^{38}$
R.~A.~Gruendl,$^{19,20}$
J.~Gschwend,$^{14,28}$
G.~Gutierrez,$^{12}$
S.~R.~Hinton,$^{39}$
D.~L.~Hollowood,$^{34}$
K.~Honscheid,$^{22,40}$
D.~J.~James,$^{41}$
A.~G.~Kim,$^{42}$
K.~Kuehn,$^{43,44}$
N.~Kuropatkin,$^{12}$
O.~Lahav,$^{16}$
M.~Lima,$^{45,14}$
H.~Lin,$^{12}$
M.~A.~G.~Maia,$^{14,28}$
M.~March,$^{46}$
F.~Menanteau,$^{19,20}$
R.~Miquel,$^{47,21}$
R.~Morgan,$^{48}$
A.~Palmese,$^{12,13}$
F.~Paz-Chinch\'{o}n,$^{19,49}$
A.~Pieres,$^{14,28}$
A.~A.~Plazas~Malag\'on,$^{50}$
E.~Sanchez,$^{31}$
V.~Scarpine,$^{12}$
S.~Serrano,$^{51,52}$
I.~Sevilla-Noarbe,$^{31}$
M.~Smith,$^{53}$
M.~Soares-Santos,$^{29}$
E.~Suchyta,$^{54}$
M.~E.~C.~Swanson,$^{19}$
G.~Tarle,$^{29}$
C.~To,$^{55,17,18}$
T.~N.~Varga,$^{56,38}$
and J.~Weller$^{56,38}$
\begin{center} (DES Collaboration) \end{center}
}}

\begin{document}
\voffset-.6in

\date{Accepted . Received }

\pagerange{\pageref{firstpage}--\pageref{lastpage}}

\maketitle

\label{firstpage}

\begin{abstract}
Strongly lensed quadruply imaged quasars (quads) are extraordinary objects. They are very rare in the sky -- only a few tens are known to date -- and yet they provide unique information about a wide range of topics, including the expansion history and the composition of the Universe, the distribution of stars and dark matter in galaxies, the host galaxies of quasars, and the stellar initial mass function. Finding them in astronomical images is a classic ``needle in a haystack'' problem, as they are outnumbered by other (contaminant) sources by many orders of magnitude. To solve this problem, we develop state-of-the-art deep learning methods and train them on realistic simulated quads based on real images of galaxies taken from the Dark Energy Survey, with realistic source and deflector models, including the chromatic effects of microlensing. The performance of the best methods on a mixture of simulated and real objects is excellent, yielding area under the receiver operating curve in the range 0.86 to 0.89. Recall is close to 100\% down to total magnitude $i\sim21$ indicating high completeness, while precision declines from 85\% to 70\% in the range $i\sim17-21$.
The methods are extremely fast: training on 2 million samples takes 20 hours on a GPU machine, and 10$^8$ multi-band cutouts can be evaluated per GPU-hour. The speed and performance of the method pave the way to apply it to large samples of astronomical sources, bypassing the need for photometric pre-selection that is likely to be a major cause of incompleteness in current samples of known quads. 
\end{abstract}
\begin{keywords}
gravitational lensing: strong --
methods: statistical --
astronomical data bases: catalogs
\end{keywords}

\section{Introduction}
\label{sect:intro}

Strong gravitational lenses are extremely valuable sources of information about the Universe. They provide unique information about the expansion rate of the Universe, the properties of distant sources that would be too faint (compact) to be detected (resolved), and about the distribution of mass in the Universe \citep[][and references therein]{Treu:2010}. 
Unfortunately, they are very rare on the sky, because the phenomenon requires the almost perfect alignment of a background source with a foreground deflector. 

Quadruply imaged quasars are a very special case of strong lensing. They are especially valuable because of the wealth of information they provide, including, for example, three independent time delays and flux ratios. At the same time, they are especially rare because they require an intrinsically rare source (quasar) to be within the inner caustic of a foreground massive galaxy. Based on the model by \citet{O+M10}, the density of quads in the sky is expected to be of order 10$^{-2}$ deg$^{-2}$ with total flux brighter than $i\sim20$ (i.e. $\sim400$ in the full sky), but only a fraction of those will be resolved and identifiable in ground based wide-field imaging of the kind obtained by the Dark Energy Survey \citep[DES, e.g.,][]{Treu:2018}. Even though the numbers have improved considerably in the past few years, only $\sim$60 quads are known across the entire sky at the time of this writing.

There are two main challenges in identifying quads from ground based optical imaging data. The first challenge is the sheer volume of data one has to inspect, considering that there are about $\sim 10^8$ stars and galaxies in the sky brighter than $i\sim20$ \citep{Annis:2014}. The second challenge is that many of the quads are only partially resolved in ground based images and thus difficult to identify and separate from astronomical contaminants. In order to overcome the first challenge, many search teams rely on color preselection to reduce the number of astronomical sources. The second challenge then becomes more manageable with a combination of algorithms applied to the image pixels and visual inspection. In the end, even in the most successful cases, confirmation via spectroscopy or higher resolution imaging (from space or from the ground with adaptive optics) is needed. Considering the cost of spectroscopy and high resolution imaging, most searches so far have focused on obtaining high purity candidate lists with high confirmation rates \citep[e.g.][]{Lemon:2020}. The drawback of this process is that many lenses are lost along the way, as evidenced by the low completeness of searches so far.

We present a new machine learning based approach to finding quadruply imaged quasars. Machine learning techniques have been applied with success to lens searches before \citep{AgneKelly15,WAT17,Hezaveh_fast_2017,Petrillo_KDS_2017,Petrillo_2018_KDS,CMU_deeplens_2017,Pourrahmani_lensflow_2018,Schaefer_2018,Avestruz_automated_2019,Madireddy_modular_2019,Cheng_unsupervised_2020,Jacobs:2019,Jacobs_candidatesDES_2019}. While building on the work in this area, our effort differs in two main ways. First, we focus exclusively on quadruply imaged quasars, developing a realistic training set using real astronomical images from the Dark Energy Survey coupled with macro and millilensing models. Second, we avoid any need for image preselection with the goal of running our algorithm on complete flux-limited samples, which in turns requires our method to be extremely fast. This may allow us to recover quads that would have otherwise been lost in preselection steps, while retaining sufficient purity to be cost-effective for follow up. To achieve these goals we apply several new techniques and methodological improvements over previous astrophysical work, such as polar convolutions, the use of multiple networks, and attention masking. In addition to focusing exclusively on quads, tests on validation datasets suggest our method outperforms machine learners used for previous searches for lensed quasars \citep{AgneKelly15,WAT17}.

In this first paper of a series we describe the method, the training set, and the results on validating datasets. A follow-up paper will present the results on a search on actual Dark Energy Survey data. The paper is organized as follows. Section~\ref{sec:methods} provides some background on the machine learning methods. Section~\ref{sec:training} describes the training set. Section~\ref{sec:methodology} describes the machine learning methods used. Section~\ref{sec:results} describes the application of our machine learning algorithms to the problem and evaluates the performance on validating datasets. A summary is given in Section~\ref{sec:summary}. 

\section{Elements of Deep Learning}
\label{sec:methods}

In recent decades, machine learning, and particularly deep learning, have demonstrated extraordinary success in tackling a wide range of tasks related to computer vision and natural language processing, benefiting fields ranging from healthcare to the development of self-driving cars, among many others \citep{lecun_DL_deeplearning_Nature_2015, wang_DL_perspective_review_2016}. In this section, We briefly review the basic elements of deep learning that we use in our models.
A more thorough introduction to deep learning can be found in \citet{Goodfellow_DL_BOOK_2016}.

Deep learning builds on simple artificial neural network (ANN) models dating back to the perceptron algorithm~\citep{rosenblatt_perceptron_1958}. 
Loosely inspired by biological neural networks, ANNs employ computational units referred to as neurons. A single neuron receives a set of inputs, represented by vector $X$, either directly from the stimulus (data), or from the outputs of a preceding set of neurons. Each neuron takes a weighted sum of its inputs, $\langle w,X \rangle$, adds a offset (aka ``intercept'' or ``bias'') term $b$, then passes this weighted sum through a function $g$ to arrive at that neuron's activation level, $g(\langle w,X \rangle + b)$. 
The term $b$ is often absorbed into the weight vector by simply appending a constant 1 to the dimensions of $X$ and letting $b$ be the corresponding weight, allowing the activation level to be written more simply 
as $g(\langle w,X \rangle)$. When $g(\cdot)$ is chosen to be non-linear, this allows for more complicated models to be built than those represented by conventional (linear) models, particularly as layers are added to the neural network.

%

\begin{figure}
\centering
\includegraphics[width=3.2in]{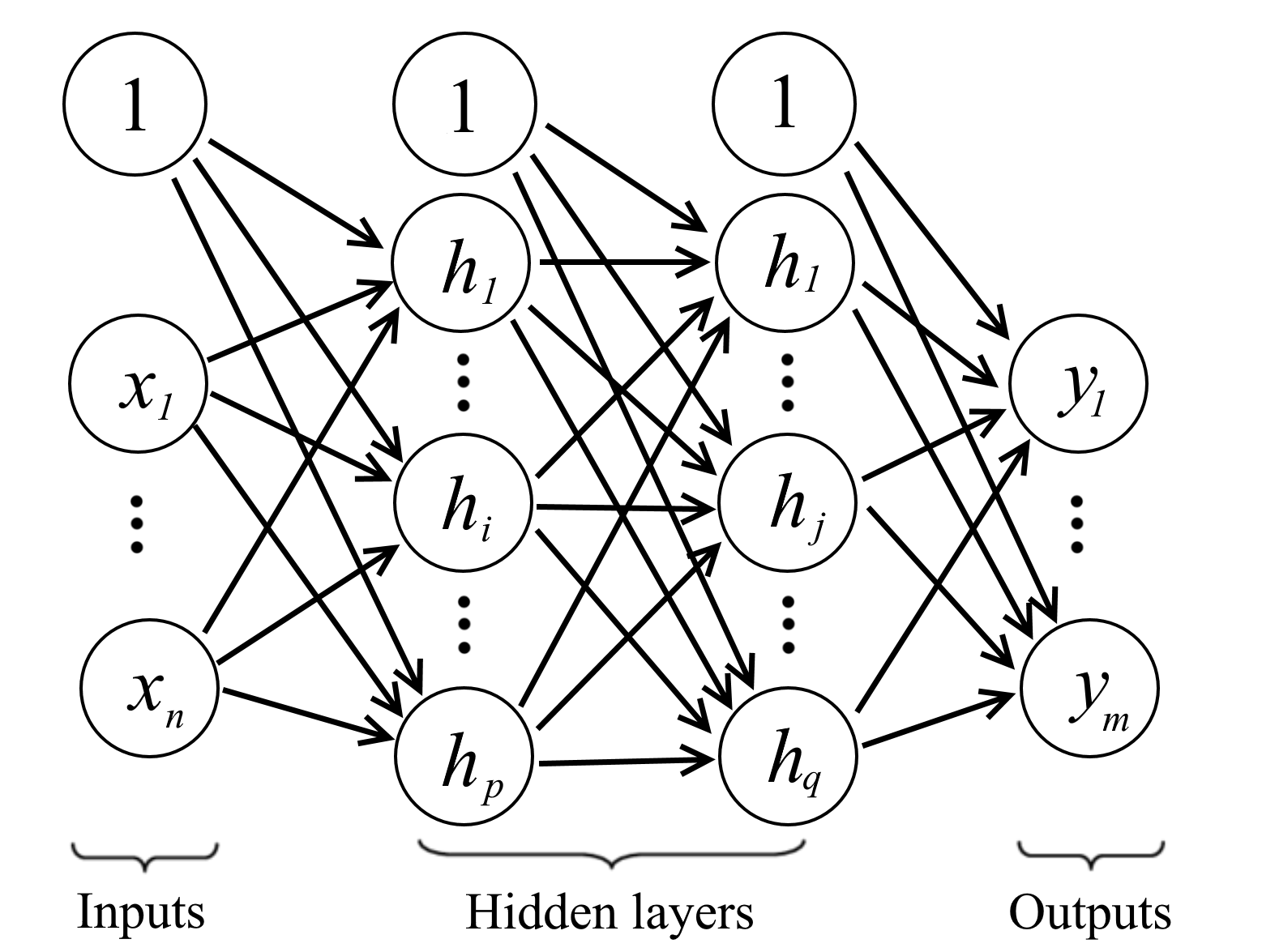}
\caption{An example of a feed-forward neural network comprised of the input layer, two hidden layers, and the output layer. The information flows in one direction from each layer to the next, and the connections are called \emph{dense} since every neuron in one layer is connected to every neuron in the next one.}
\label{fig:dl_MLP}
\end{figure}

The arrangement of neurons in layers is specified by an \emph{architecture} that, for each group of neurons (called layers), determines from which other groups they receive their inputs and to which groups they send their outputs. Layers in which every neuron is connected to every neuron of the next layer are called \say{dense} (see Figure~\ref{fig:dl_MLP}).
The first layer of neurons are activated directly by the data. 
The final layer is the output layer that generates the quantity of interest, for example, the predicted class of the input in a classification task.
An ANN with more than a few layers is often called \emph{deep}, 
in contrast to early-generation ANNs that typically employed only one or two \emph{hidden} layers located between the input and output layers. Adding layers increases the learning capacity of the network, but exacerbates the difficulties of fitting or ``training'' the model. 

\subsection{Training the model} Neural networks are trained by adjusting the weights so that they optimally convert input data $X$ (e.g. an astronomical image) into the desired output $Y$ (e.g. a prediction of whether the image contains a lensed quasar). During the training, the ANN takes input data $X$ and generates an output, $\hat{Y}$, for each training example, by propagating computations forward through the network: each neuron computes $g(\langle w,X \rangle)$ using the inputs ($X$) given to it and the current value of the weights (inclusive of the bias term), $w$. The difference between the generated output, $\hat{Y}$, and the correct answer, $Y$, is assessed using a loss function, $\mathcal{L}(Y, \hat{Y})$, chosen according to the nature of the problem. Various algorithms are available to determine how the weights of the network should be adjusted in light of each error. Typically these approaches assess how each weight contributes to the error, in a process that works backwards through the model computing the gradient of the loss using the chain rule. The weights can then be updated by some fraction of the value needed to correct the response, a procedure generally referred to as backpropagation as it propagates the error signal backwards through the network. Numerous optimization methods have been proposed with varying performance. One popular choice, used here, is the adaptive momentum (Adam) algorithm \citep{Kingma_DL_Adam_2015}, which computes individual, adaptive learning rates for the model parameters considering the first and second moments of the gradients.



Training is usually iterative, with each step using a portion of the training data (called a \say{mini-batch}) and adjusting the weights according to learning rates that control the speed of convergence. The end goal of the optimization is to reach a global minimum, a set of values of the parameters where the loss function is minimized. However, as the number of parameters is typically very large (in the millions for all models here), we generally expect multiple local minima. Stochastic gradient descent avoids getting ``stuck'' in sub-optimal local minima by randomly selecting mini-batches. 

Because these models are so flexible, they can lead to \say{overfitting}, where the model learns specific characteristics of the training dataset associated with particular outcomes, but which do not generalize well to unseen data. This results in large errors when the model is applied to new data. Numerous methods help to prevent this. 
One important approach is to add terms to the loss function that effectively penalize models with more extreme weights. This constrains the model and avoids results that depend heavily on very specific features but that might have turned out very differently in different samples. Such an penalization scheme is known as regularization and can be represented as seeking to minimize 

\begin{equation}
\begin{aligned}
\mathcal{L}^*(Y, \hat{Y}) = \mathcal{L}(Y, \hat{Y}) + \lambda \Omega(W),
\end{aligned}
\label{eqn:param_regularization}
\end{equation}
where the scalar $\lambda$ controls the strength of regularization and $\Omega$ defines a regularizing functional. In the past, the first and second norms were frequently chosen as $\Omega$ and referred as $L_1$ and $L_2$ regularization respectively. Later, orthogonality of the weights was argued to be a desirable property since multiplication by an orthogonal matrix leaves the norm of the input unchanged. This led to orthogonal regularization, $\Omega(W) = \norm{\langle W, W\tran \rangle - I }_1$, where $I$ is the identity matrix.

\subsection{Choice of the activation function}
In many practical problems the model must be made to fit a non-linear function between the inputs and outputs, calling for a non-linear choice of activation function. 
Until recent years, the logistic (or sigmoid) function $g(x)=e^x / (1+e^x)$, hyperbolic tangent $\tanh(x) = \frac{e^x - e^{-x}}{e^x + e^{-x}}$, softmax $[s(x)]_i = \frac{e^{x_i}}{\sum_k e^{x_k}}$, and linear rectifier $\text{ReLU}(x) = \max(0, x)$ were among the most popular activation functions. However, deeper models require a choice of activation function that protects against the risk of having an error gradient equal to zero for many weights. To this end, 
functions such as parametric ReLU (PReLU) \citep{He_PRELU_ICCV_2015}, exponential linear unit (ELU) \citep{Clevert_ELU_ICLR_2016}, scaled exponential linear unit (SELU) \citep{Klambauer_SELU_NIPS2017}, and Swish \citep{swish_Ramachandran_ICLR_2018} have become widely used. These are defined as: 
\begin{equation}
\begin{aligned}
\text{PReLU}(x) &=
\begin{cases}
x, &\text{if } x \ge 0,\\
ax, &\text{if } x < 0, 
\end{cases} \; (a < 1), \\
\text{ELU}(x) &=
\begin{cases}
x, &\text{if } x \ge 0,\\
a(e^x-1), &\text{if } x < 0,
\end{cases}\\
\text{SELU}(x) &= \beta \cdot \text{ELU}(x)\text{, }\; \beta > 1, \\
\text{Swish}(x) &= x \cdot 
\sigmoid(x)
= \frac{x}{1 + e^{-x}}.
\end{aligned}
\label{eqn:dl_activations}
\end{equation}

\subsection{Convolutional neural networks} 
\begin{figure}
\centering
\includegraphics[width=3.2in]{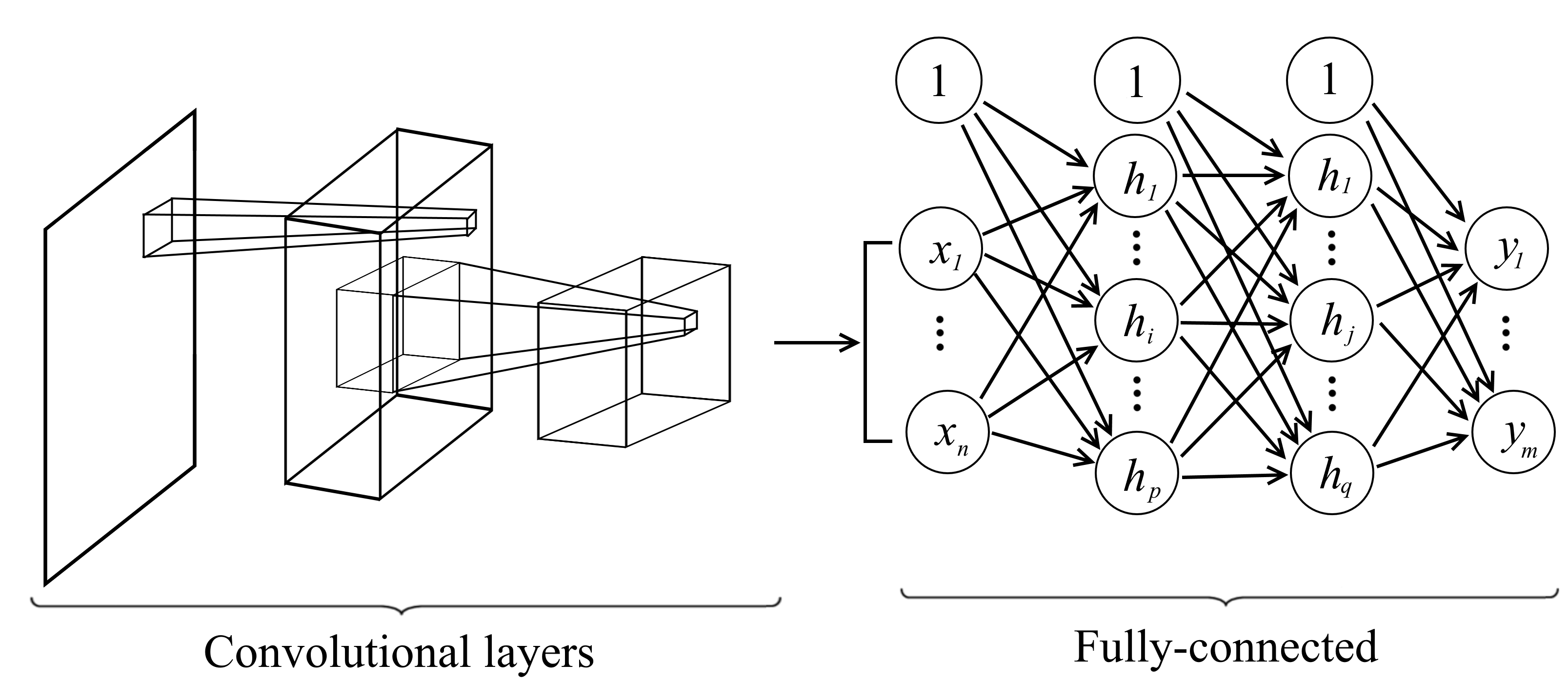}
\caption{A convolutional neural network consisting of two convolutional layers and three dense layers. The first convolutional layer produces a feature map by sliding a convolution window over the input image. Within each window the layer takes a weighted sum of the corresponding pixels to produce a single output pixel. The output feature map of the second convolutional layer is flattened into a one-dimensional vector, used as an input the following fully-connected ANN.}
\label{fig:dl_CNN}
\end{figure}

While dense ANNs with at least one hidden layer and an appropriate activation function can approximate any function, in practice ANNs can be made far more powerful, with less training data and fewer parameters to tune when they can be designed to extract the more relevant and informative features from the data. To this end, convolutional neural networks (CNNs) (see Figure~\ref{fig:dl_CNN}) have proven extremely powerful in image processing applications. 

These networks have one or more layers that, like layers of visual cortex, contain neurons whose activation's summarize key features in designated patches of the input image. 
Specifically, neurons in a given layer are receptive to a particular area of the input/image, known as their ``receptive fields''. The convolution or weighted average they perform over their receptive fields summarizes the information in it. The weights used in this convolution, which constitute a kernel or filter, can be preset or learned during training. In CNNs with multiple convolutional layers, each layer takes the previous layer's activation as its input, with wider kernels in subsequent layers, so that the receptive field corresponding to a neuron grows to cover larger receptive fields over the image.

Stacking these convolutional layers, therefore, enables learning representations of the data at different levels of abstraction. This architecture has made deep CNNs very effective in computer vision, sometimes achieving superhuman performance in classification and discrimination tasks. Units within a given layer typically share the same kernels (weights) to reduce the number of learned parameters. Non-trainable convolutional layers are often referred as pooling layers and can be used to apply basic operations such as $\max(\cdot)$ or $\text{mean}(\cdot)$ that corresponds to \say{MaxPooling} and \say{AvgPooling} layers respectively. 

\subsection{Blocks}
Blocks of neurons can be designed and connected together to develop powerful network architectures. Here we discuss three types of blocks that have proven valuable in related computer vision problems and that we consider in our own architecture.

\begin{figure*}
\centering
\includegraphics[width=17cm]{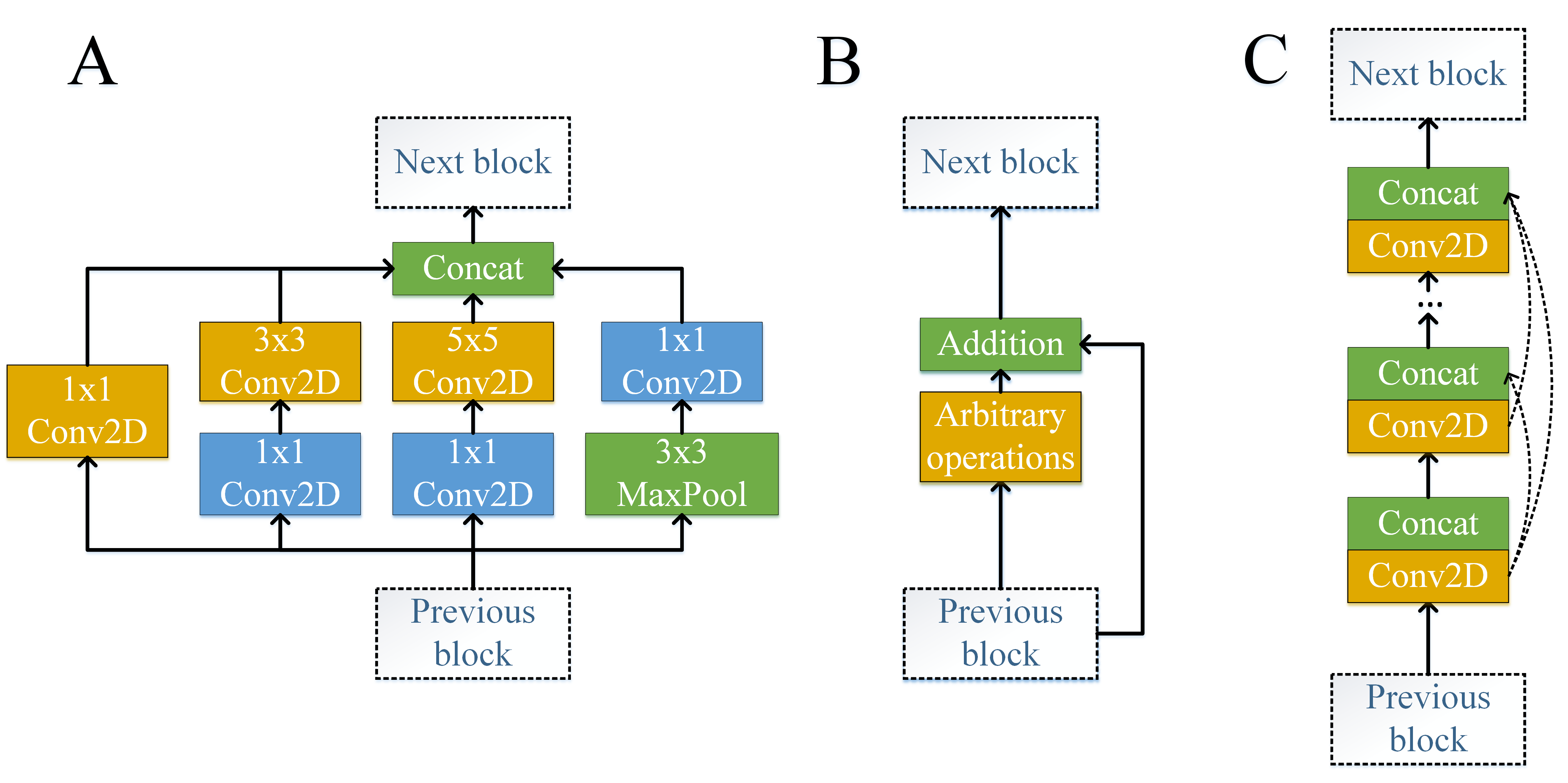}
\caption{Common computer vision building blocks: (A) inception block, (B) residual block, and (C) dense block. The inception block applies multiple convolutional operations of different window size (shown in orange) and a pooling operation (shown in green) in parallel, instead of being restricted to a single window size, and then concatenates the extracted features together. As concatenation leads to a quickly growing size of the output tensor, 1-by-1 convolutional layers (shown in blue) are used to reduce the dimensionality. The residual block employs a \emph{skip-connection} that adds the input tensor to the output tensor, which mitigates the ``vanishing gradient'' problem. The dense block similarly uses skip-connections, but concatenates the tensors instead of adding them.}
\label{fig:dl_inverse_famous_models}
\end{figure*}

The ``inception'' block, is best known from InceptionNet (also known as GoogLeNet) \citep{Szegedy_DL_GoogleNet_2015}, and later ResNet \citep{szegedy_DL_GoogleNet_v4_2017}. This block is described in Figure~\ref{fig:dl_inverse_famous_models}~A). It concatenates four versions of the processed input data, each processed in parallel: the original image, two convolutional layers that use a sequence of increasingly large receptive fields, and one simpler local-averaged/smoothed version of the image.

The next block type is the residual block. A key problem with deeper networks---and part of the reason they did not emerge in earlier decades of ANNs---is the ``vanishing gradient'' problem: the updates to the weights computed by backpropogation factor in the gradient at each step, and a long sequence of such factors produces update values close to zero.
One way to address this, employed in the ResNet architecture \citep{He_DL_ResNet_2016} is an ``identity shortcut connection'', also known as a residual or skip-connection. In this configuration, the input of each learning block is added to the output before propagating to the next one (see Figure~\ref{fig:dl_inverse_famous_models}~B). This makes it easier to propagate information forward and backward without significant alterations and simplifies training of deep models.

The third block type we consider is the dense block in the convolutional setting, as in the DenseNet architecture \citep{Huang_DL_DNET_2017}. While ResNet and its residual block use element-wise addition, dense convolutional blocks combine processed inputs of one layer (here, a convolutional layer) with lower level data by concatenation instead of addition. Each layer thus receives feature maps from all preceding convolutions, within the same block (see Figure~\ref{fig:dl_inverse_famous_models}~C). Later blocks may then use pooling or other approaches to reduce dimensionality. DenseNet has become widely used in various computer vision problem such as image classification, object detection, and image segmentation due to superior computational efficiency and quality of the learned features. 

\subsection{Generative modeling and data representation}
\begin{figure*}
\centering
\includegraphics[width=17.8cm]{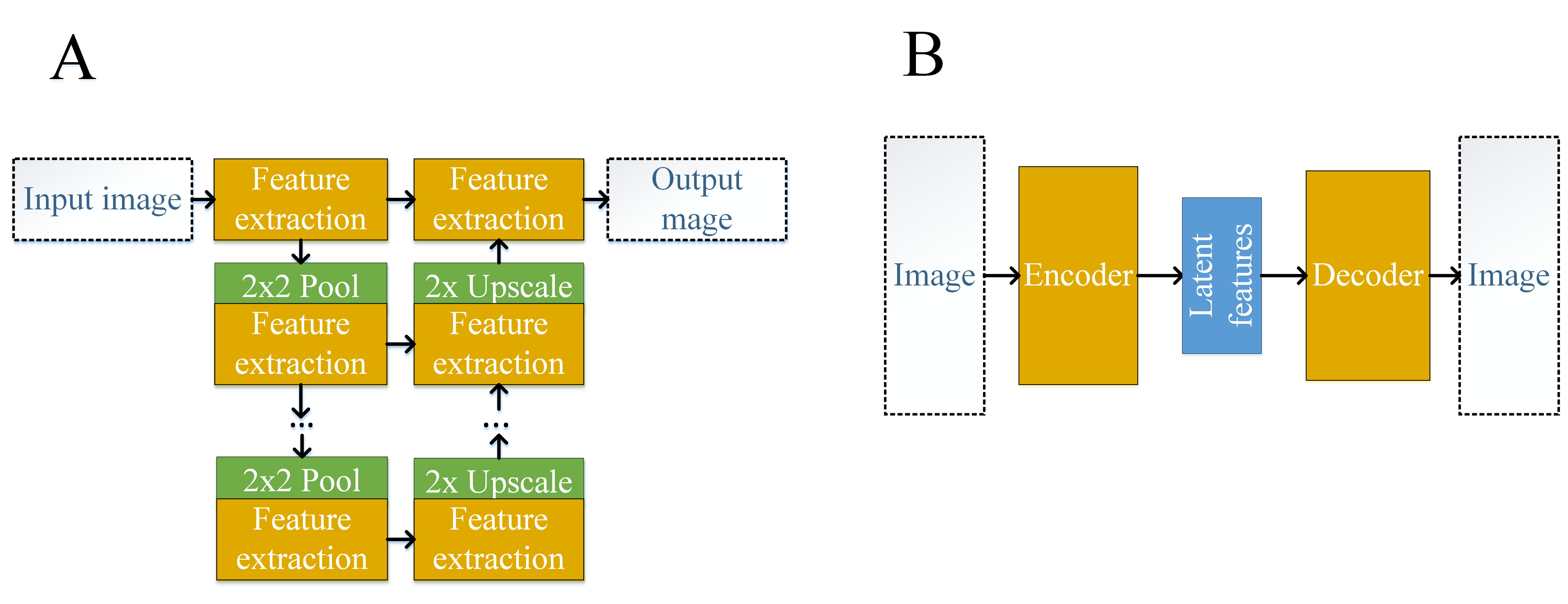}
\caption{Data representation and generative modeling: (A) U-Net and (B) autoencoder. The U-Net model contains a contracting path and an expansive path that down-scales and up-scales the input image to allow simultaneous feature extraction at higher and lower image resolutions. The autoencoder model has two major parts: an encoder that maps the input image to the latent features, and a decoder that maps the latent features to a reconstruction of the input image. As the dimensionality of the latent features is often smaller than the dimensionality of the input image, the model learns a compressed representation of the data and removes noisy components.}
\label{fig:dl_inverse_generative_net}
\end{figure*}

Some tasks in computer vision are ``image-to-image'' problems, in which we have one input image and desire to create another related image of the same size. These including denoising (removing artifacts from an input image), segmentation (estimating a set of binary masks that encode different regions on the input image), and detection (locating objects on the input image) tasks. 
The U-Net is one important architecture for such problems. The model was first popularized in biomedical image segmentation \citep{Ronneberger_DL_UNET_2015}. In this context it takes input images (e.g. CT scans) and outputs segmentation masks that show regions of interest (e.g. malignant tumors), based on information from a labeled dataset. The U-net architecture is so named because it contains contracting and then expansive paths (see Figure~\ref{fig:dl_inverse_generative_net}~A). The contracting path employs a series of feature extraction blocks followed by one or more ``scaling down'' layers, such as a pooling layer. The expansive path concatenates the features of the same resolution, fuses the extracted features, and up-scales the image representation. The up-scaling method could be anything from a non-trainable nearest-neighbor interpolation to a trainable deep CNN. Additional skip connections between individual blocks of the contracting and expansive pathways enable easier gradient propagation. Importantly, the symmetry between two paths and the U-shaped architectures lets the network propagate context information to higher resolution layers. 

Another valuable class of image-to-image architectures is the autoencoder (AE), which operate on unlabeled data and provide efficient, latent space representations. The architecture has two major parts connected sequentially: encoder and decoder (see Figure~\ref{fig:dl_inverse_generative_net}~B). The learning objective is to reconstruct the original image as effectively as possible, as judged by a loss function, while forcing the information to pass through a lower-dimensional middle block, labeled ``latent features'' above. The activations of neurons in this layer thus offer a lower dimensional representation of the input image, sufficient to recover the best reconstructed image possible. It is thus valuable as an unsupervised means of learning the important features of images.

One limitation of the lower dimensional representations learned by AE is that they can follow an arbitrary distribution, which may lead to situations where samples of the same class have drastically different latent features. Variational AE (VAE) is often used to alleviate this deficiency \citep{vae_tutorial}. VAE differs from AE in two ways: its latent features are selected pseudo-randomly and its loss function is extended by a penalty term. During training, each latent feature is independently drawn from a Gaussian distribution $\mathcal{N}(\mu, \sigma)$, where the parameters $\mu$ and $\sigma$ are taken from the output vector of the encoder $\{\mu, \log{(\sigma})\}$. This exposes the decoder to a range of encoding vectors (as opposed to a single vector in AE) forcing it to map neighbouring feature vectors to the same image. The loss function is penalized by a Kullback--Leibler divergence of the latent features for a given input sample and the standard normal distribution. The penalty term ensures that the encoder refrains from producing extreme values and encourages it to evenly distribute around the center of the latent space. This leads to a continuous and orthogonal latent space, a highly desired property for data representation.

\section{Methodology: Generation of Training datasets}
\label{sec:training}
\begin{figure*}
    \includegraphics[scale=1.2]{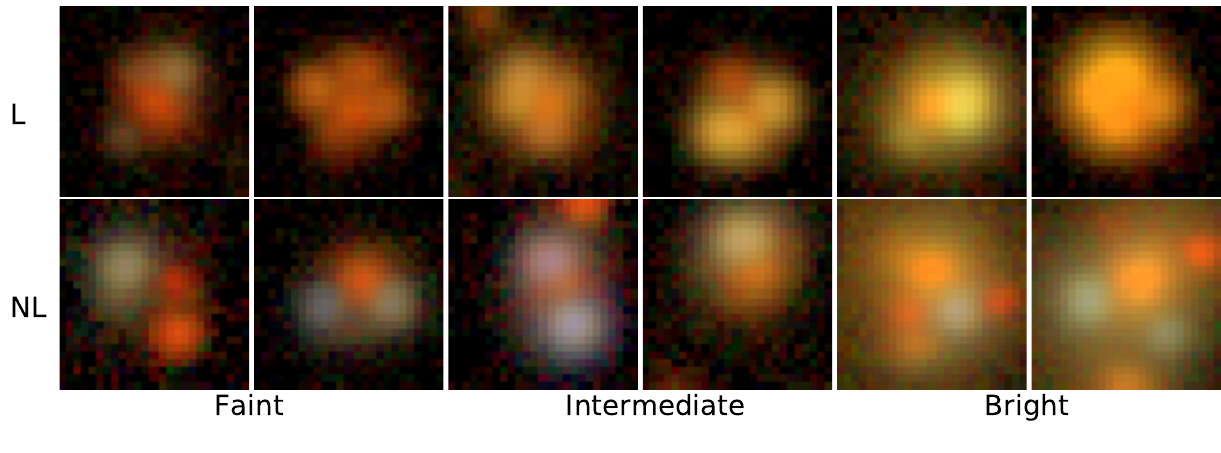}
    \caption{Training sample comprising of lenses (top) and non-lenses (bottom). The left-most two columns show objects selected from the fainter sample ($i>19.0$), middle two columns show objects from the intermediate sample ($18.5<i<18$) and the right-most two columns show the objects from the brighter sample ($i<17.5$). The images are 6.75~arcsec on the side.}
    \label{fig:trainsamp}
\end{figure*}

We generate a training sample of lensed quasars and known sets of contaminants for training of the network. Since only a few tens of true lensed quasars have been confirmed, we cannot use them alone to construct a training set, and must generate simulated lensed quasars based on their well understood physics. Ensuring the realism of these simulated observations is essential to the ultimate generalization of our model to real data. We use a version of {\sc simct} \citep{More:2016} modified for this purpose. Leaving the details to \citet{More:2016}, we begin by using the redMaGiC galaxy catalog \citep{Rozo_etal_redMaGic_2016} as our parent galaxy sample. All galaxies from this sample are considered potential lenses. By using the measured redshift and magnitudes and known scaling relations, we estimate the lens mass assuming that the mass density profile follows a singular isothermal ellipsoid. We assume mass follows light to determine the centroid, ellipticity and position angle of the lens. We also include external shear to account for effects due to objects in the immediate environments of the lens galaxy. We draw sources from known luminosity functions with a certain i-band magnitude and redshift. Colors are then extracted from the quasar catalog of \citet{Tie2017} by cross-matching the source i-band magnitude and redshift. Given the lens parameters and source parameters, we calculate the lensing cross-section and determine if a source would be lensed by the potential lensing galaxy such that the multiple images can be well-resolved and above the limiting magnitude.

We further implemented the microlensing magnification effect by stars within the lensing galaxy which can affect the fluxes of the lensed quasars. For a given lens and source, we calculate the positions and fluxes of the lensed quasar images. The microlensing effect increases or decreases the flux of the lensed images as determined by the local convergence ($\kappa$), shear ($\gamma$) and smooth matter fraction ($s=1-\kappa_{\ast}/\kappa_{\rm tot}$) as described, e.g., by \citet{Vernardos:2019}. In order to optimize computing resources, we compute microlensing magnification maps for a large number of fixed values of $\kappa$, $\gamma$ and $s$ and interpolate from this grid to real cases. Stars are assumed to have masses 1 M$\odot$ and the stellar density profile is assumed to follow the de Vaucouleurs profile \citep{deV48}. We determine the convergence due to compact (stellar) population $\kappa_\ast$ in the image plane following \citet{Vernardos:2019}. The resulting sample consists of about 28500 simulated lensed images of background quasar which are then added on top of the redmagic galaxies from the DES-Y3 data \citep{Sevilla-Noarbe2021}.
We show a few simulated lenses in the top row of Figure~\ref{fig:trainsamp} with fainter systems on the left end and brighter on the right. There are two examples for each of the faint ($i>19.0$), intermediate ($18.5<i<18.$) and bright ($i<17.5$) magnitude bins. The image cutouts are 6.75~arcsec on the side where each pixel is 0.27~arcsec wide matching the DES pixel resolution. 

For the training set of non-lenses, we use the spectroscopically confirmed stars from the Sloan Digital Sky Survey data and photometrically selected quasars \citep{Tie2017} and blue cloud galaxies \citep{Williams2017}. As we are interested in quads, we randomly draw objects from these catalogs and place them randomly around a massive galaxy which could mimic a lensed quad. About 2000 such systems are generated and this sample size is increased by a factor of five by applying rotations. We show examples of these non-lenses in the bottom row of Figure~\ref{fig:trainsamp}. As before, fainter systems are on the left end and brighter on the right. These examples correspond to the same magnitude bins as the simulated lenses in the top row. As part of the non-lens sample, we also include the same redMaGiC galaxies that were used to generate simulated lenses but without any lensing features around them. This resulted in 28,500 of simulated positive examples (lenses) and 28,500 of negative examples (contaminant galaxies).

\begin{table*}
    \centering
    \caption{Details of various samples used in our analysis.}
    \begin{tabular}{l|l|c|c}
        \hline
         Sample Name  & Description & \multicolumn{2}{c}{Sample size}  \\
                  & & Original & Augmented \\
        \hline
        Simulated lenses & RedMagic deflectors + simulated lensed quasars & 28,500 &    1,254,000 \\
        Simulated non-lenses & RedMagic deflectors + contaminants & 2,000 &  128,000 \\
        RedMaGiC non-lenses & RedMagic galaxies & 28,500 & 1,140,000  \\ 
        \hline
        Augmented data & Augmented simulated lenses and non-lenses, and RedMaGiC non-lenses & - & 2,522,000 \\
        Training dataset & 40\% of the augmented data & - & 1,008,800 \\
        Validation dataset & 40\% of the augmented data & - & 1,008,800 \\
        \hline
        Test dataset & 20\% of the augmented data & - & 504,400 \\
        Confirmed real systems & Spectroscopically confirmed true positives and negatives from STRIDES & 119 & 128 \\
        Mixed dataset & Random samples from the test dataset and additionally augmented confirmed real systems & - & 1216 \\  
    \end{tabular}
    \label{tab:my_label}
\end{table*}

\section{Methodology: Model architecture and training}
\label{sec:methodology}




Our modeling methodology involves several steps. First, we pre-process, augment, and split the data for purposes of model training and tuning. Then, we employ unsupervised learning methods to explore the data and aid in feature extraction. Next, we train a series of supervised learning models with a variety of architectures. Finally, we develop an ensemble over the resulting models. In this section we describe each of these steps in turn.
\subsection{Data pre-processing and splitting}

We first standardize each image so that the pixels in each image (pooled together across all four griz-bands) have zero mean and unit standard deviation. More specifically, let $I_{ipg}$ denote the intensity in griz-band $g \in \{1,2,3,4\}$ of pixel $p$ in image $i$. Let $\mu_i$ and $\sigma_i$ be the sample mean and standard deviation of $\{I_{ipg}\}_{p, g}$, respectively. We normalize each pixel by computing 
\begin{align}\label{eq:inst:norm}
(I_{igp} - \mu_i) / (\sigma_i + \varepsilon)
\end{align}
across all $i,g$ and $p$, where $\varepsilon > 0$ is a small perturbation introduced for numerical stability. We refer to this procedure as \emph{instance normalization}.



The resulting images are augmented with random rotations to ensure a rotation invariant result. We then split the data into training, validation, and testing subsets. 
This produces roughly $10^6$ images in each of the training and validation sets and $5 \times 10^5$ in the test set. 

The models described below are trained on the training set, with hyperparameters optimized by minimizing the prediction loss on the validation dataset. After training and choosing hyperparameters, the training and validation data are combined to be re-used for training. The resulting model is then evaluated for its performance on the testing set.

\subsection{Unsupervised learning}\label{sec:unsup}

\begin{figure*}
\centering
\includegraphics[width=6.4in]{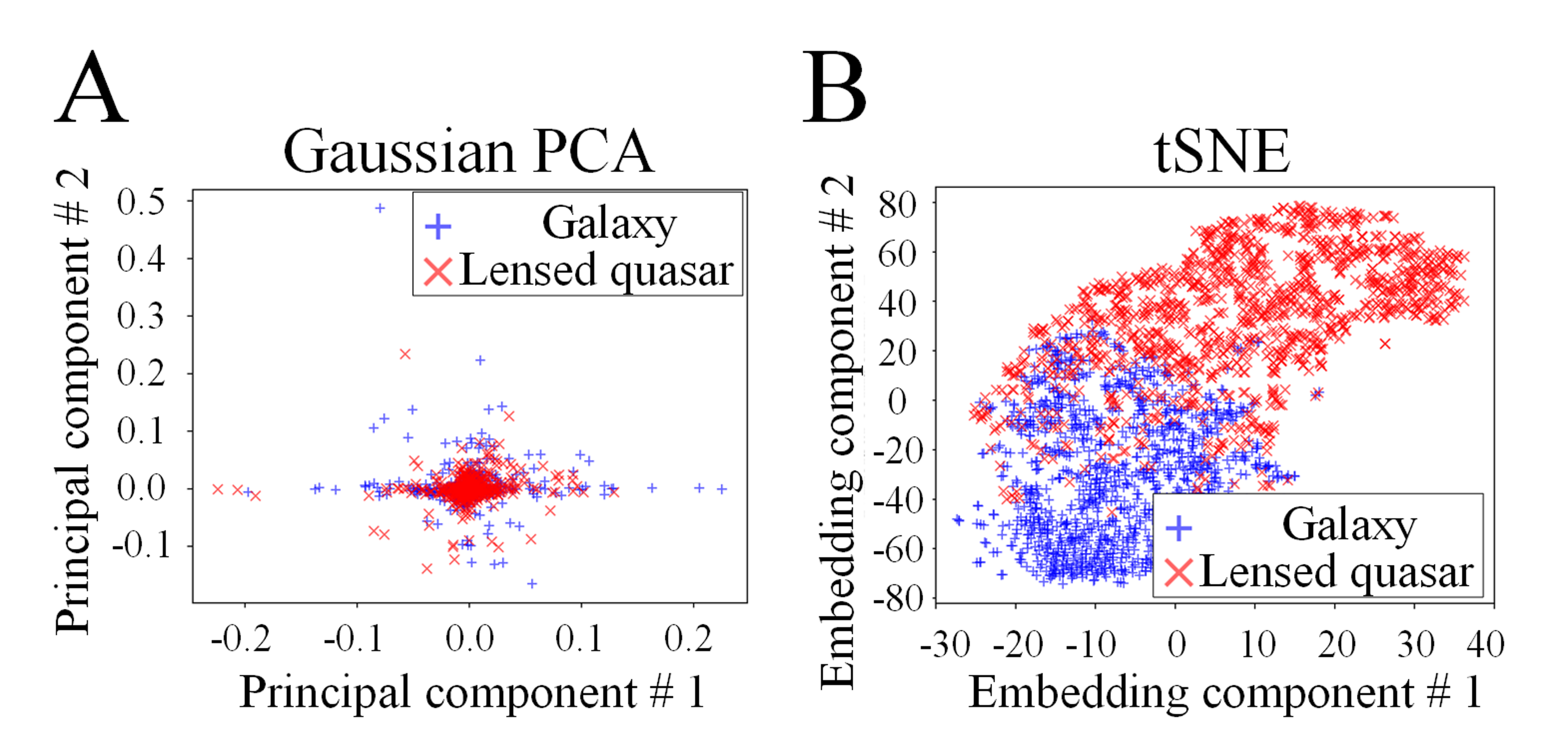}
\caption{Visualization of a subset of the data consisting of galaxies (blue) and lensed quasars (red) after dimensionality reduction: (A) the first two components of the Gaussian Principal Component Analysis (PCA) and (B) two-dimensional t-distributed stochastic neighbor embedding (tSNE). PCA shows little evidence of the separability of lenses from non-lenses. The two-dimensional tSNE, however, suggest that the classes are reasonably separable in a higher dimensional space.}
\label{fig:dim_red}
\end{figure*}

To aid in constructing features that would facilitate the supervised learning process, we begin with an unsupervised exploratory process on a subset of $100,000$ images sampled from augmented data. We first decompose the dataset using a Gaussian kernel principal components analysis (kPCA). Explaining 95\% of the variance in the dataset requires $8$ components. Plotting just the first two principal components (Figure~\ref{fig:dim_red}A) shows little evidence of the separability of these two classes. Fortunately, however, the higher-dimensional information does suggest the classes are reasonably separable. This is visualized using the two-dimensional t-distributed stochastic neighbor embedding (tSNE) of the images in Figure~\ref{fig:dim_red}B, which uses the relative similarity of lens images with each other as compared to non-lens images to infer the possibility of successfully distinguishing these objects in a sufficiently rich non-linear feature space.


\begin{figure*}
\centering
\includegraphics[width=5in]{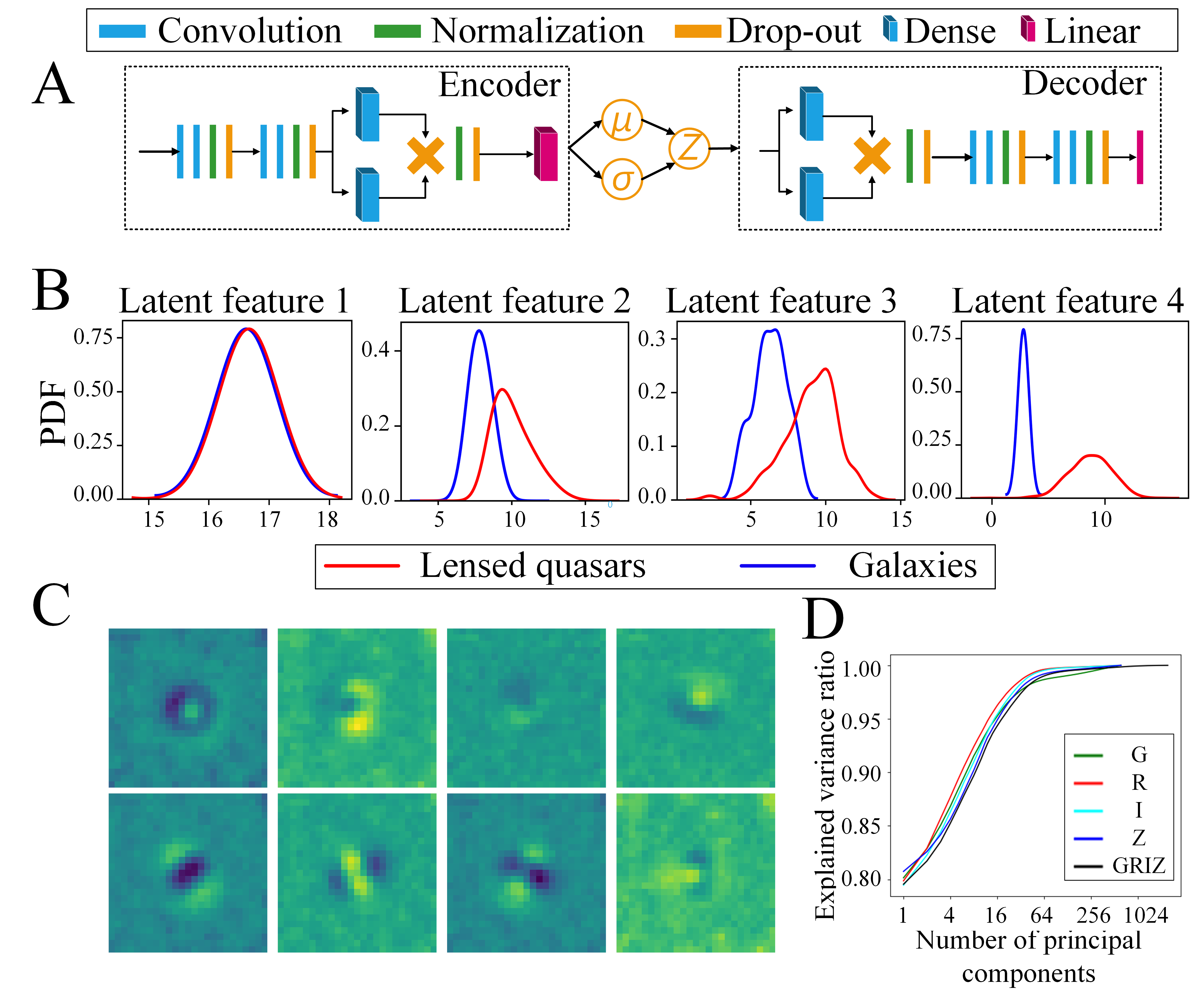}
\caption{Unsupervised exploration of the latent space of the dataset: (A) a VAE model used for learning the mapping from images to the means and standard deviations of the latent variables; (B) empirical distributions of selected latent variables showing that despite nearly identical statistical properties in galaxies and lensed quasars across the majority of learned latent variables, some of them describe features that can be used to distinguish between these two classes; (C) a visualization of a few learned features of the lower dense layer of the VAE model generated by putting the learned weights into square shapes of the same size as the output of the previous convolutional layer; (D) empirical relationship between the explained variance and the number of principal components of the latent variables of 1024-dimensional VAE suggests that a 128-dimensional latent space can be sufficient to describe the dataset.}
\label{fig:latent_space}
\end{figure*}

Next, Figure~\ref{fig:latent_space}A depicts the architecture of VAE which we use to generate an orthogonal, lower-dimensional latent space characterization. The encoding CNN (left side of the VAE) contains two blocks, each with two convolutional layers followed by normalization and dropout layers. The purpose of the first convolutional block is to extract low-level features using $5 \times 5$ kernels. The second convolutional block further expands the effective receptive field using $9 \times 9$ kernels to extract higher order features. Note that this configuration makes the effective receptive field of the last convolutional layer of $25 \times 25$, covering the entire input image. The drop-out layer is used to reduce overfitting by dropping a portion of the randomly selected inputs.

The subsequent block consists of two parallel dense layers that process the input tensor and map it to the latent space of dimension $p$. The upper layer applies the sigmoid activation to its output to estimate a \textit{feature-selecting vector} $\sigmoid\bigl(W_{upper} z\bigr)$ for each training sample. By multiplying it with the latent space representation of an image $z$ produced by the output of the lower layer, $W_{lower} z$, we locally suppress irrelevant latent features of each sample:
\[ 
(W_{lower} z) \odot \sigmoid\bigl(W_{upper} z\bigr),
\] where $W_{lower}$ and $W_{upper}$ are $p \times q$ matrices and $z$ is a flattened image of size $q \times 1$. The output dense layer, shown in pink in Figure~\ref{fig:latent_space}A, linearly combines the remaining features to estimate the means and standard deviations. Finally, as shown on the right half, the process is repeated in reverse to reconstruct an image from the compressed features. 

To explore the result, Figure~\ref{fig:latent_space}B shows how the two classes are distributed along four selected latent features. As VAE gives a mean ($\mu$) and a standard deviation ($\sigma$) for each latent feature, the aforementioned figure depicts empirical distribution of the means of the latent features. We observed that although many features such as feature 1 do not discriminate contaminant galaxies from lensed quasars, some of the features such as feature 4 do separate the two classes well. To better understand the meaning of the latent features, we plot some of the weights of the lower dense layer in Figure~\ref{fig:latent_space}C. The plots depict the underlying modes that naturally span the dataset including those that prevail among lensed quasars. To see how dimensionality of the latent space impacts the explained variance along individual bands (g, r, i, z) or their combination (griz), we apply linear PCA on the features of $1024$-dimensional VAE models that were trained to reconstruct a single band or the entire image. The results in Figure~\ref{fig:latent_space}~D suggest that $16$-dimensional space covers $95\%$ of the variance in griz-bands (around $12$ latent variables for R), while $128$ features correspond to $99\%$ of the variance (around $64$ latent features in R). This shows that VAE models can learn meaningful features and describe the dataset with a $128$ latent variables. We later employ an encoding portion of the VAE models for feature extraction and subsequent image classification.

\subsection{Supervised learning}
This section provides details for the construction and fitting of several neural network architectures, before describing how they are combined in an ensemble.

The unsupervised learning explored in Section~\ref{sec:unsup} indicates that lensed quasars have distinctive geometric features that could be utilized in their detection. We frame lens detection as a binary classification problem in which \say{one} corresponds to a sample with a lensed quasar and \say{zero} corresponds to anything else. A classifier takes an image as an input and outputs a binary label. In practice, the output layer has two units, whose activations we can refer to as $z_0$ and $z_1$. To these we apply the softmax function to obtain quantities we interpret as probabilities, i.e., $\hat{y}_1 = \pr(Y=\text{``lens''})=\exp(z_1)/(\exp(z_0)+\exp(z_1))$, with $\hat{y}_0 = \pr(Y=\text{``not lens''}) = 1 - \pr(Y=\text{``lens''})$. We found empirically that training with the softmax activation results in a better generalization compared to training with the sigmoid activation.

For the loss function, we employ the binary cross-entropy, also known as the negative log-likelihood, between a class label $y \in Y$ and the predicted probabilities $(\hat{y}_1,\hat{y}_0)$,
\[
\mathcal{L}(y, \hat{y}_1, \hat{y}_0) = -y\cdot \log{(\hat{y}_1)} - (1-y)\cdot \log{(\hat{y}_0)}.
\] Moreover, binary cross-entropy leads to a more consistent gradient propagation as the 
log-terms mitigate
exponential behavior due to gradient saturation for extreme values.

Instead of the conventional batch normalization, we use instance normalization introduced in~\eqref{eq:inst:norm}.
While instance normalization makes the samples of both classes statistically indistinguishable in terms of the pixelwise mean and variance, it has several important advantages. First, it limits the range of the values that tensors can take, preventing saturation of hyperbolic activation functions, and improving gradient propagation. Second, it improves generalization by avoiding overfitting on synthetic data, as the gap between the simulated and the real objects is often the major issue in problems that rely on statistical models trained on synthetically-generated data. In particular, it reduces the effect of the outliers that could cause significant covariate shift in hidden layers of the ANN.

To penalize overfitting, we extensively use dropout layers. The exact order of the hidden layers, activation functions, normalization layer, and dropout layers is typically optimized for every problem. We empirically found that the combination of activation layers followed by normalization and dropout layers achieve a better bias--variance trade-off. To additionally limit overfitting, we introduce zero-mean Gaussian noise at the input of each model, which adds random noise to the input images at every training step. We set the  noise covariance matrix to 
$\sigma^2 I$,
where $I$ is the identity matrix and  $\sigma^2 = 0.062$, with the value determined by hyper-parameter optimization. Note that this is equivalent to additional data augmentation performed simultaneously with training.

Models are trained using the Adam algorithm \citep{Kingma_DL_Adam_2015}. Adam uses adaptive learning rates for every model parameter, enabling faster convergence. We also employ the scheduled (staircase exponential) and triggered decrease of the learning rate. The latter changes the step size when the loss rate of decay is below a certain threshold.

To further combat overfitting, we employ early stopping based on the validation loss. More specifically, after every iteration on the training set, we evaluate the loss function on the validation set which includes data not used for optimizing model parameters.
We stop once the validation, rather than the training, loss has converged.

Besides trainable parameters, each model has hyperparameters, a small set of values that define model architecture and training dynamics. These include model depth (number of hidden layers), model width (number of convolutional filters and dense units), model resolution (size of convolutional kernels), and the choice of the activation function, regularization strength, and learning rate. We use the recently proposed Hyperband algorithm \citep{hyperband}, an efficient bandit-based method for hyperparameter optimization. It allocates computational resources to as many configurations as possible and throws out those that show poor performance over time until a single configuration remains. This method maximizes the number of tested configurations and results in a more efficient resource utilization compared to the grid search, random search, or Bayesian optimization. 

\subsubsection{Prior methods}

Numerous prior works proposed using the traditional computer vision models such as InceptionNet, ResNet, and DenseNet for lens searches. These architectures have demonstrated exceptional performance on traditional computer vision problems, but the vast number of trainable parameters they require can result in either poor generalization (overfitting) or weak convergence (underfitting). Pre-training these models on the ImageNet, a large dataset of ordinary images collected for object recognition, degrades the performance further. We hypothesize that this is due to drastic differences in visual features between the lensed quasars and ImageNet samples.


Fortunately, several earlier projects have sought to detect lenses using various deep CNN architectures. We describe these here before proposing two novel architectures. Previously proposed CNN models include CMU DeepLens \citep{CMU_deeplens_2017}, LensFlow \citep{Pourrahmani_lensflow_2018}, CNNS \citep{Jacobs_candidatesDES_2019}, and several others \citep{Hezaveh_fast_2017, Schaefer_2018, Avestruz_automated_2019}. 
We reproduce and retrain LensFlow and CNNS on our data for comparison. LensFlow is based on a classical architecture comprised of three convolutional layers, maximum pooling layer and four dense layers all connected in series. Drop-out layers in-between the dense layers mitigate over-fitting. Unusually, LensFlow applies hyperbolic tangent to the outputs of the convolutional layers and ReLu activation function to the output of the dense layers. 
CNNS combines four convolutional layers into blocks of two layers each with three subsequent dense layers. Each convolutional block is followed by pooling operations, halving the dimensionality of the feature maps the output. The activation function is ReLu, and drop-outs are again used to mitigate over-fitting. 

\subsubsection{NaiveNet}
\begin{figure}
\centering
\includegraphics[width=3.2in]{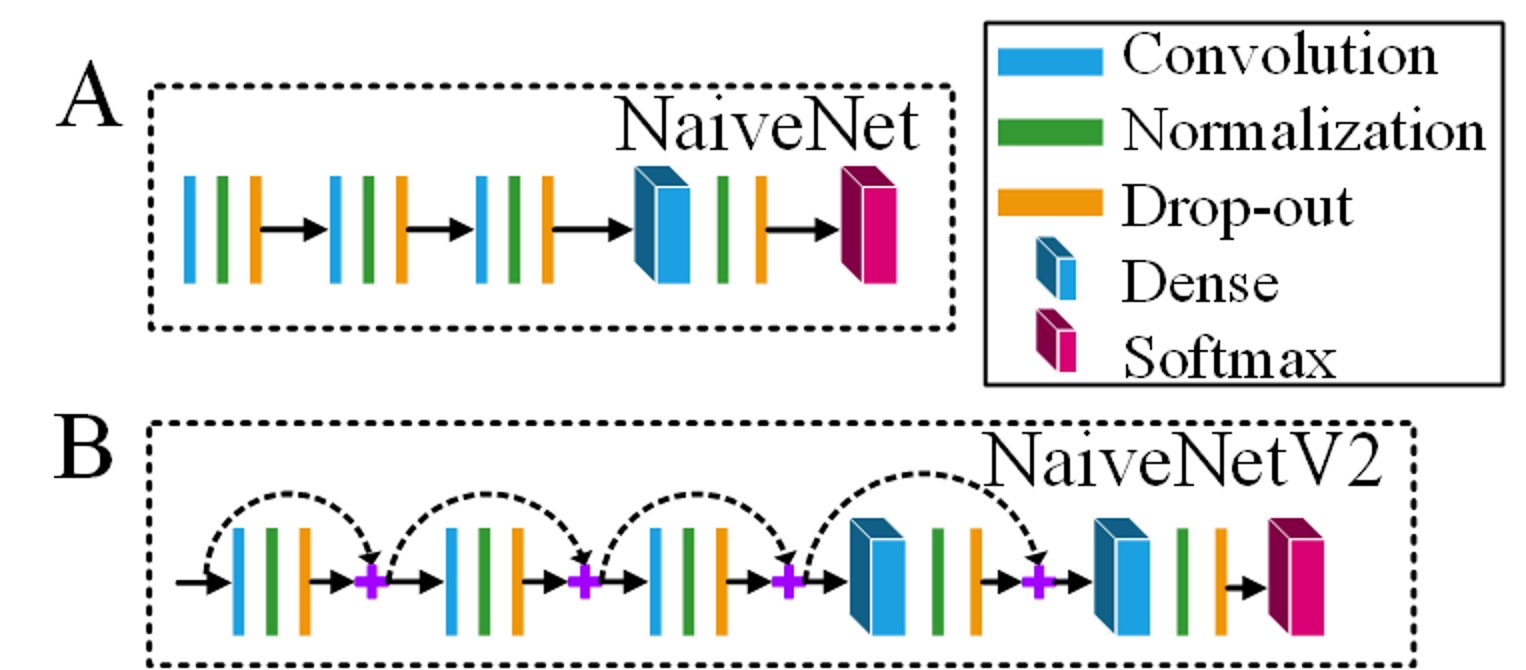}
\caption{Optimized instances of generalized CNN architectures: (A) NaiveNet based on three convolutional blocks and two dense layers and (B) NaiveNetV2 that uses residual connections and adds an additional dense layer to improve the performance.}
\label{fig:naive_cnn}
\end{figure}

Inspired by these early examples, we first propose a CNN involving several convolutional layers followed by dense layers, which we refer to as NaiveNet, illustrated in Figure~\ref{fig:naive_cnn}. The number of layers, number of dense units (or convolutional kernels) within each layer, and size of the convolutional kernels respectively, are optimized together with other hyperparameters. We intentionally avoid specific constraints on the architecture and, instead, optimize performance through a hyperparameter search. Next, we introduce NaiveNetV2, replacing convolutional and dense blocks with their residual alternatives. Each residual path has two consecutive layers: the first one with a chosen non-linearity and the second one with a linear activation function.

\subsubsection{Polar convolution}
\begin{figure}
\centering
 \includegraphics[width=3.2in]{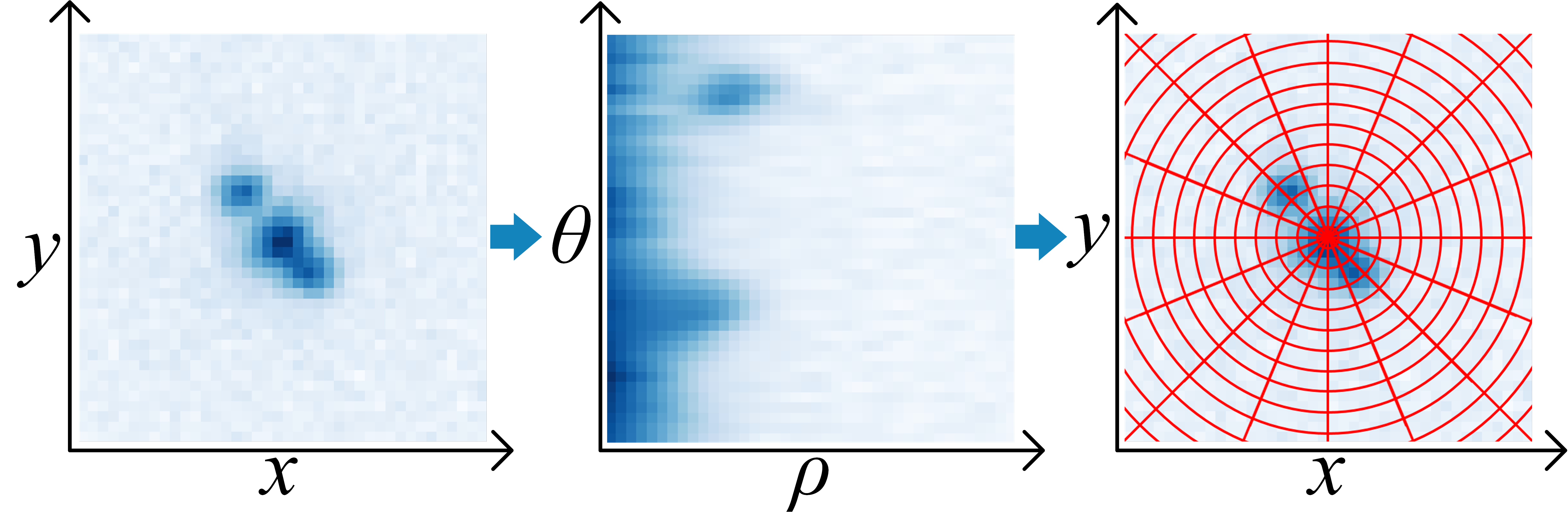}
\caption{Visualization of the proposed polar convolution operation: each feature map gets converted to higher resolution polar coordinates where a regular rectangular convolution is applied and then translated back to Cartesian coordinates. Effectively, the operation has a convolutional window of an angular sector of annuli of different radius (shown in red) with smaller window size at the center of the image.}
\label{fig:polar_conv}
\end{figure}

Next, we propose an alternative approach to feature extraction. Since we can ensure positioning of the objects at the center of the griz images and given that the gravitational field originates from the center of mass of the objects, a polar coordinate system is a natural choice. We define the corresponding 2D polar convolution similarly to its rectangular counterpart:

\begin{equation}
  \label{eqn:polar_conv}
  \big(W * T(z)\big)_{\rho, \theta} = \sum_{i}\sum_{j} W_{i,j} \cdot T(z)_{\rho + i, \theta+j}
\end{equation}
where $T(\cdot)$ translates an input tensor from Cartesian to a polar coordinate system. It first transforms a fixed size rectangular tensor $z$ into a scatter field of points $z^*$ with polar coordinates $\{\rho,\theta\} = \{\sqrt{(x^2+y^2)}, \arctan{(y/x)}\}$. Then, it estimates a smooth rectangular tensor $s$ using biharmonic spline interpolation:

\begin{equation}
  \label{eqn:spline}
  \begin{cases}
  s_i = \sum_j \alpha_j g(i, j) \\
  g(i, j)=\norm{p_i - p_j}^2 (\log{\norm{p_i - p_j}} -1)
  \end{cases}
\end{equation}
where $g$ is Green's function, $p_i = [x_i, y_i]$ is a vector pointing at the corresponding scatter point with value $z^*_i$, and the weight vector $\alpha$ can be found by solving $g \cdot \alpha = z^*$. An illustration of the proposed operation is shown in Figure~\ref{fig:polar_conv}. Beyond fitting the underlying geometrical structure on the images, polar convolution yields a heterogeneous receptive field size and a particularly sparse connectivity matrix by focusing at the central part of the image, which is useful as the corners of the images are expected to be less informative, with more noise. 

\subsubsection{Visual attention masking}
\begin{figure}
\centering
\includegraphics[width=3.2in]{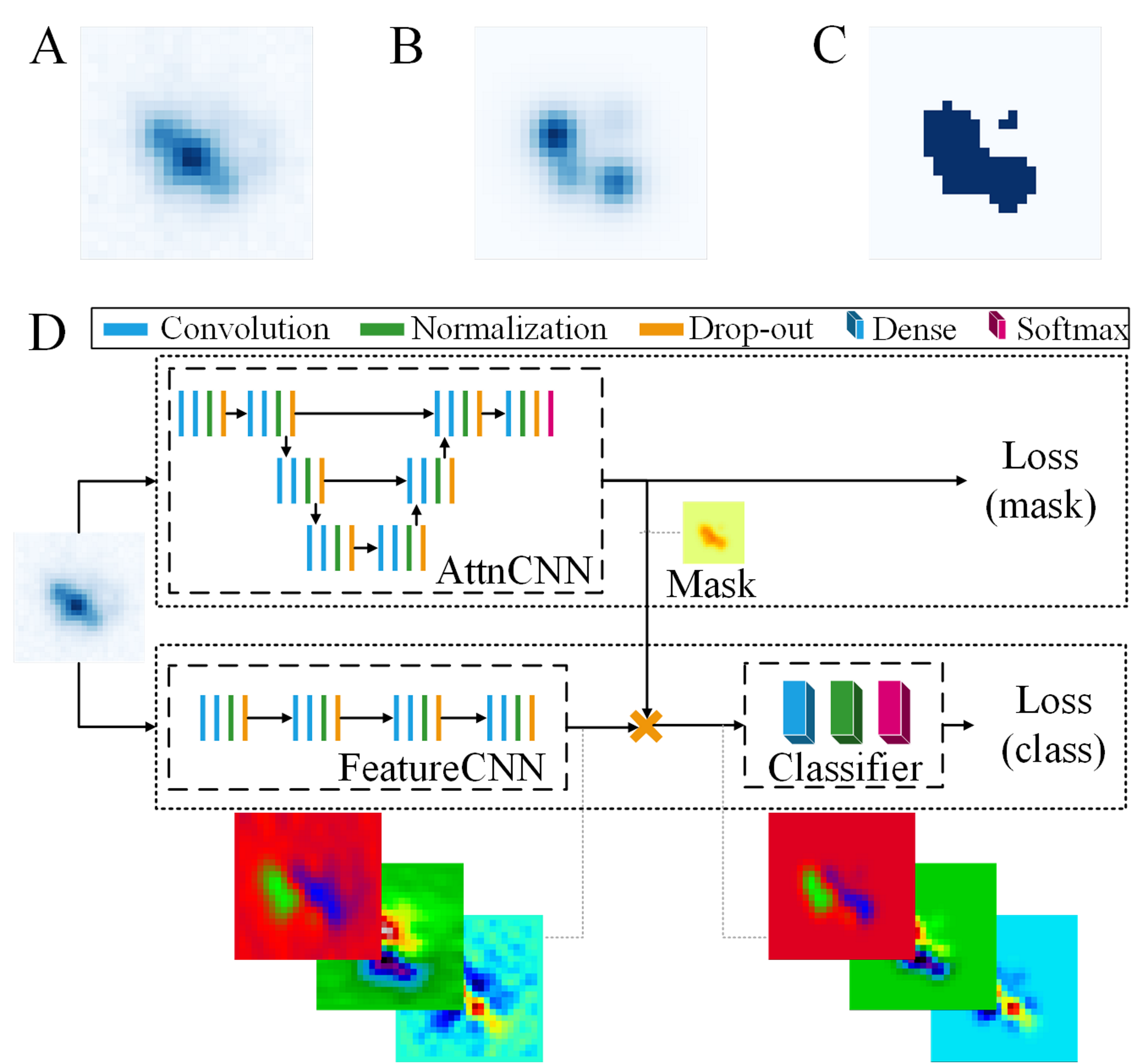}
\caption{An example of the proposed attention masking for feature extraction: (A) a sample of the original input image; (B) the corresponding lensed quasar blended in the original image; (C) a binary mask used as a label for training the segmentation model. The mask is produced by setting a threshold on intensity of the lensed quasar image. (D) The proposed AttnNet architecture is comprised of three models: the AttnCNN trained to segment regions containing lensed quasars, the FeatureCNN that extracts features from the original image and then applies the estimated binary mask to remove noisy components, and a binary classifier trained to identify lensed quasars.}
\label{fig:attn_cnn}
\end{figure}

We train an additional U-Net-like model that produces a binary mask representing regions of lensed quasars (named as AttnCNN in Figure~\ref{fig:attn_cnn}~D). The goal of this auxiliary masking network is to guide feature extraction by bringing attention to regions likely to have important features. We found empirically that thresholding image pixels with 22.5\% of their maximum intensity creates meaningful binary masks as the brightest regions are the most informative. We tried several loss functions for training the AttnCNN and found that mean absolute error leads to the most stable results. It was beneficial to train the entire assembly simultaneously in an end-to-end fashion. At each optimization step $i$ (batch size of $n$), we update the parameters of the AttnCNN model $\Psi(\cdot)$ to minimize the discrepancy between its outputs $\Psi(X_i)$ and the binary masks $M_i$: 

\[ \mathcal{L}_i^{\text{AttnCNN}}=\frac{1}{n}\sum_{k=1}^{n}\abs{\Psi(X_i)_k - M_{ik}}.\] At the same time, we independently update parameters of the feature extracting model $\Lambda(\cdot)$ (named as FeatureCNN in the figure) and the binary classification model $\Pi(\cdot)$ by backpropagating the gradient of the cross-entropy loss ${\mathcal{L}_i\Big(\Pi\big[\Psi(X_i)\odot\Lambda(X_i)\big], Y_i \Big)}$.
This way AttnCNN directly affects the performance of the feature extracting CNN and the classifier, which forces FeatureCNN to adjust its learned parameters.

\subsection{Argus: ensemble of models}
One of the challenges we encountered was overfitting stemming from the gap between the simulated and real data. Particularly, we observed a significantly lower performance during testing on real samples generated from the DES, despite the imposed regularization and optimal synthetic testing results. Given the limited size of the real dataset, we found it impractical to directly bridge the gap by finding a projection between real and synthetic samples. Moreover, since the confirmed objects do not represent the entire range of possible lenses, additional training on them would further amplify the bias. 

\begin{figure*}
\centering
\includegraphics[width=6.4in]{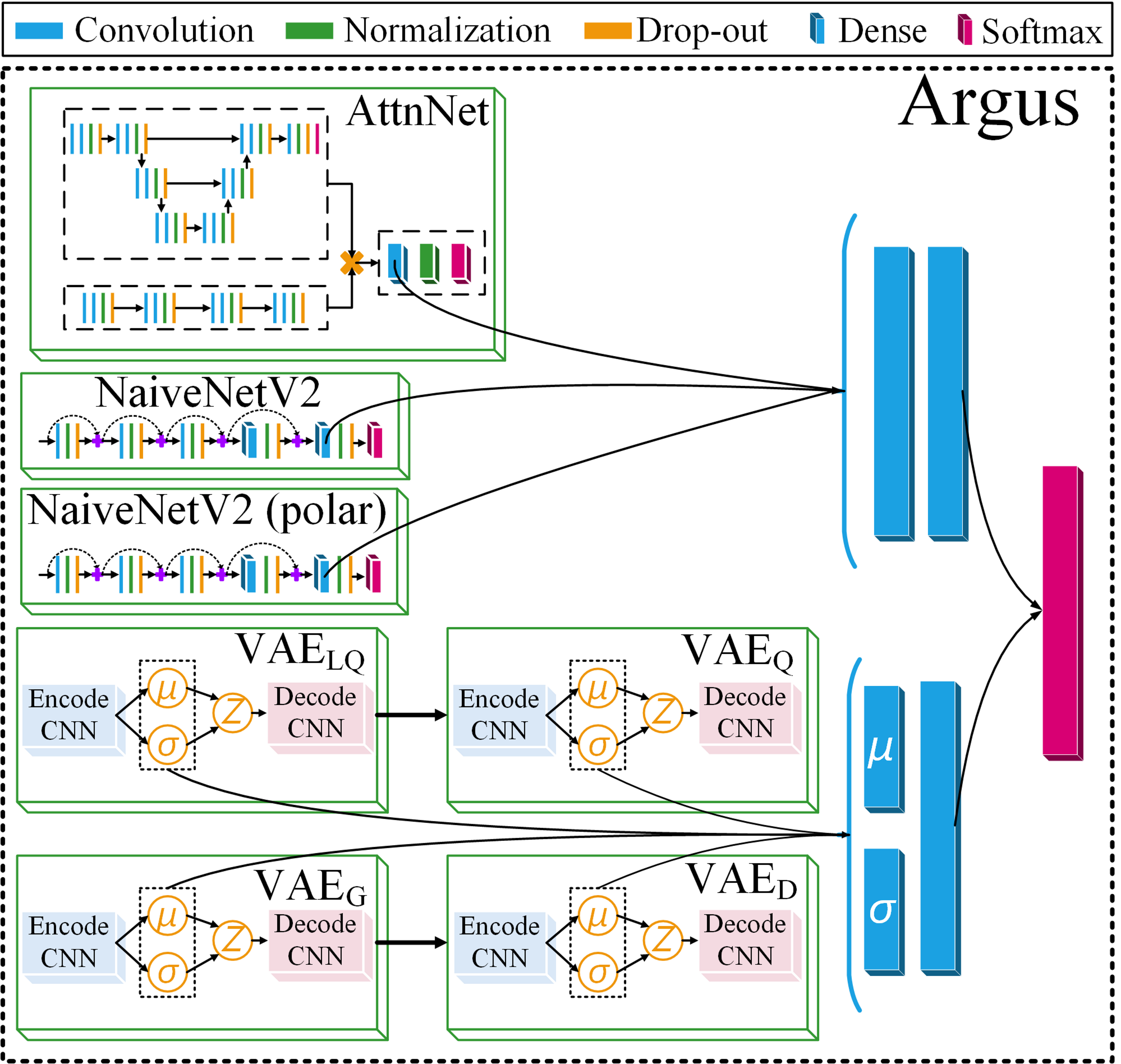}
\caption{Argus: an end-to-end ensemble of deep neural networks. The proposed method includes feature extraction elements of pre-trained models: a 512-dimensional output from the first dense layer of AttnNet, and a 128-dimensional vector from the second dense layer of NaiveNetV2. These features are concatenated together and then reduced to 128-dimensional space through two dense layers. In addition, the model takes 128-dimensional latent features of four serially-connected VAE models trained on reconstructing images of lensed quasars (LQ), galaxies (G), delensed quasars (Q), and deflector fields (D) from the blended input image. As the encoders produce both means and standard deviations of these features, they are first processed independently and then combined into a 64-dimensional feature vector. The final output is produced by a dense layer that classifies samples based on 192 features.}
\label{fig:argus_model}
\end{figure*}

Instead, a more practical and efficient method to improve generalization is model stacking (sometimes called blending), which combines the outputs of pre-trained models. We employ this technique by blending classifier models and VAEs (see Figure~\ref{fig:argus_model}). The upper branch maps the hidden feature spaces of the AttnNet and NaiveNetV2 based on rectangular and polar convolutions to another $128$-dimensional space. It employs two dense layers with ReLu non-linearity to fuse a $512$-dimensional vector from the AttnNet with a pair of $128$-dimensional vectors from the rectangular and polar NaiveNetV2: $\Phi_{s}: \mathbb{R}^{512+128+128} \mapsto \mathbb{R}^{128}$. The lower branch mixes the latent spaces of VAE models that were pre-trained to reconstruct specific components on the images including lensed quasars ($\text{VAE}_{LQ}$), contaminant galaxies ($\text{VAE}_{G}$), underlying source quasars ($\text{VAE}_{Q}$), and deflectors ($\text{VAE}_{D}$) that represent the strength of the gravitational field of the lensing galaxy. We found that reconstruction on the quasars and deflectors achieved the best results when conditioned on the output of the $\text{VAE}_{LQ}$ and $\text{VAE}_{G}$ respectively. This is implemented by stacking the second order VAE model on top of the decoder and using the mean component of the latent features of the first encoder. We choose a $128$-dimensional latent space since it is able to converge to a global minimum of the loss function and it was earlier suggested by the results of PCA to keep the explained variance ratio above $99\%$. However, because galaxies and lensed quasars have an identical distribution along the majority of the latent space axes, we apply a stronger dimensionality reduction: $\Phi_{u}: \mathbb{R}^{512+512} \mapsto \mathbb{R}^{64}$. Note that the standard deviations estimated by the encoding CNN models carry information about the noise in the images and the associated confidence levels. Since $\mu$ and $\sigma$ often have different scales and distributions we first blend them separately and then combine new intermediate feature spaces. The output layer linearly combines the features of the upper and lower branches to classify the input samples.

To train the entire end-to-end model we split the training dataset into two halves. We train each individual component (supervised and unsupervised) on the first half. Then, after stacking, we train the blending dense layers using the second half of the data. This scheme allows us to minimize additional overfitting as the blending layers learn how to combine the feature spaces of the pre-trained models on previously unseen samples. Moreover, to foster generalization we use normalization and dropout layers as well as combination of $L_1$ and $L_2$ (\say{elastic net}) regularization. The corresponding hyperparameters of the model and its components are selected using the Hyperband search described earlier.

\section{Results}
\label{sec:results}

\subsection{Training and testing setup}
After training these models we tested their performance on three datasets that were not used during the training. First, the \textit{simulated objects} is a conventional \say{testing} dataset that was sampled from the data generated for model training as described in Section~\ref{sec:training}. It contains around $250,000$ lensed quasars and $250,000$ contaminant galaxies. Second, the \textit{past candidates} dataset contains $20$ confirmed lensed quasars and $108$ confirmed non-lenses that were previously proposed as candidates for spectroscopic follow-up within the STRIDES collaboration (as collated by one of us, C.L.). Finally, the \textit{mixed dataset} contains $1216$ objects randomly sampled from the \textit{simulated objects} and \textit{past candidates}. We augmented the samples from \textit{past candidates} via random rotations and mirroring to achieve a roughly 1:1 ratio. 

We trained each model on a computer with a single GPU (Nvidia GeForce GTX 1080 Ti). A full training cycle of a single model with hyperparameter optimization took between $20$ to $180$ hours depending on the model. In addition to our proposed models, we reproduced two previously reported models, CNNS and LensFlow, retraining these on our data. In doing so we followed the design choices as reported, but adjusted learning rates and the corresponding parameters (decay rate, learning rate schedule, etc.) to maximize performance. 

\subsection{Performance}
As a first performance metric we consider the area under the receiving operating curve (AUROC). This can be interpreted as the probability that the model ranks a random positive example more highly than a random negative example.
\begin{table}
\centering
\caption{Classification performance (AUROC).}
\resizebox{0.45\textwidth}{!}{
\begin{tabular}{|l|c|c|c|}
\hline
Model \& parameters & \specialcell{Simulated\\objects} & \specialcell{Past\\candidates} & \specialcell{Mixed\\dataset} \\
\hline
CNNS (8.8M) & 0.889 & 0.580 & 0.853\\
\hline
LensFlow (8.8M) & 0.901 & 0.593 & 0.855\\
\hline
NaiveNet (10.2M) & 0.901 & 0.605 & 0.862\\
\hline
NaiveNetV2 (2.4M) & 0.914 & 0.684 & 0.863\\
\hline
NaiveNetV2 polar (2.4M) & 0.937 & 0.582 & 0.864\\
\hline
AttnNet (5.8M) & 0.940 & 0.774 & 0.878\\
\hline
Argus (17.7M) & 0.997 & 0.650 & 0.894\\
\hline
\end{tabular}}
\label{tab:testing_results}
\end{table}

Table~\ref{tab:testing_results} shows the results. Both existing models, CNNS and LensFlow, had approximately 8.8 million trainable parameters and demonstrated AUROC values of roughly 0.85 on the mixed data, comparable to our initial NaiveNet model. The enhanced version with residual skip-connection, NaiveNetV2, achieved an AUROC of 0.86 on the mixed dataset while reducing the number of parameters to 2.4 million.

Attention masking (AttnNet), however, led to a substantial improvement in performance with an AUROC of 0.940 on the simulated objects and 0.774 on the past candidates. AttnNet was the most conservative in detecting lenses and had the highest AUROC on the past candidates dataset, which is driven by the built-in feature selection and stronger orthogonal regularization. Finally, the proposed ensemble model (Argus) reached a nearly perfect AUROC score of 0.997 on the simulated data, but had a slightly lower score on the past candidates compared to AttnNet. On the mixed data, Argus achieved an AUROC of 0.894, the highest among all models.

\begin{figure}
\centering
\includegraphics[width=3.2in]{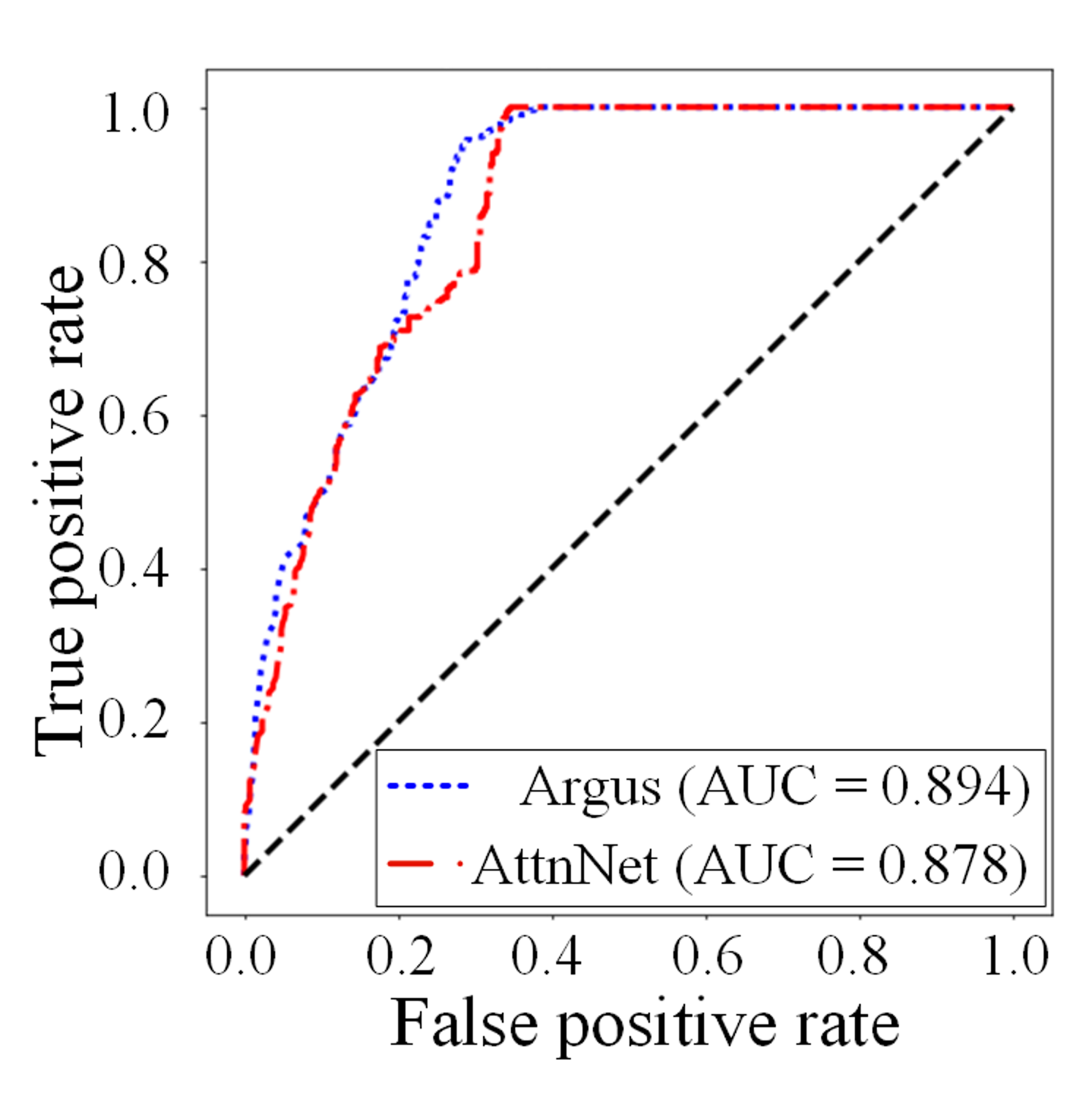}
\caption{Receiver operating characteristic (ROC) curves of the best performing models on the \textit{mixed dataset}, which show the possible trade-offs between the probability of a correct detection (true positive rate) and the probability of false detection (false positive rate). The Argus ensemble achieves a nearly perfect true positive rate while keeping the false positive rate under 0.3. The AttnNet demonstrates true positive rate of about 0.8 for the same false positive rate under 0.3.}
\label{fig:mixed_roc}
\end{figure}

ROC curves of the AttnNet and Argus models on the \say{mixed dataset} are shown in Figure~\ref{fig:mixed_roc}. The Argus ensemble achieves a nearly perfect true positive rate while keeping the false positive rate under 0.3. The AttnNet demonstrates true positive rate of about 0.8 for the same false positive rate under 0.3.

\begin{figure}
\centering
\includegraphics[width=3.3in]{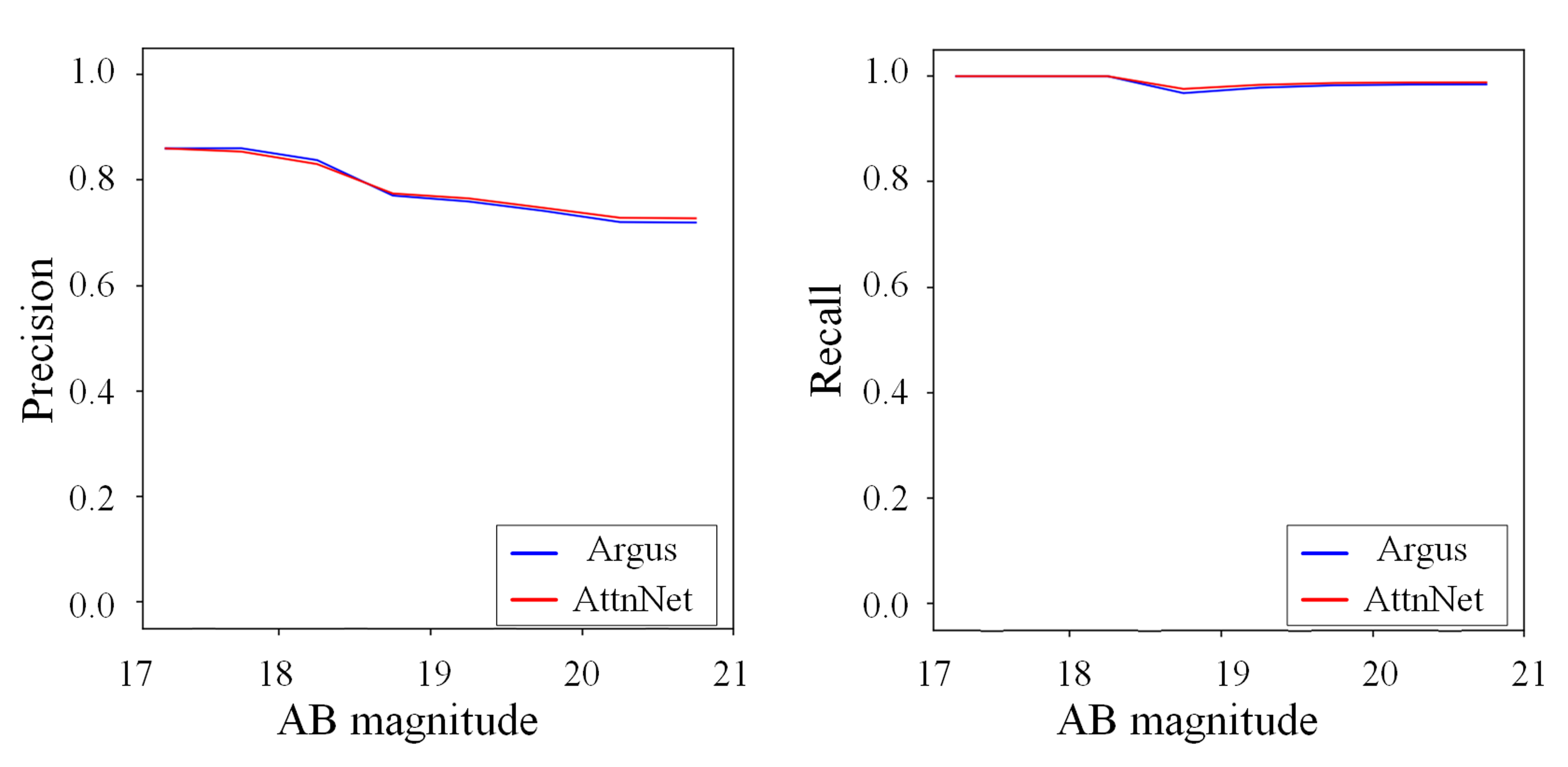}
\caption{Precision and recall as a function of the AB magnitude of the whole system (obtained by summing up the flux in the cutout) estimated on the \textit{mixed dataset}. For comparison, the 10-$\sigma$ limiting magnitude of the data for point sources is 23.34.}
\label{fig:mixed_ab_mag}
\end{figure}

Figure~\ref{fig:mixed_ab_mag} depicts how precision (the fraction of true lenses among the predicted lenses) and recall (the fraction of correctly identified lenses among true lenses) of two best performing models change with AB magnitude of the whole system (obtained by summing the flux over the entire cutout; for comparison, the 10-$\sigma$ limiting magnitude of the data for point sources is 23.34). To produce these plots we grouped the samples of the mixed dataset into five bins of equal width in AB magnitude and evaluated the metrics for each group. We found that as AB magnitude increases the proposed models start suffering from a noticeable drop in precision. This is expected because of the declining signal to noise ratio of the images and false positives become more difficult to identify as such. Remarkably, however, recall is uniformly close to 100\% suggesting that our method should achieve high completeness even when the signal to noise ratio is low.

\begin{figure*}
\centering
\includegraphics[width=17.8cm]{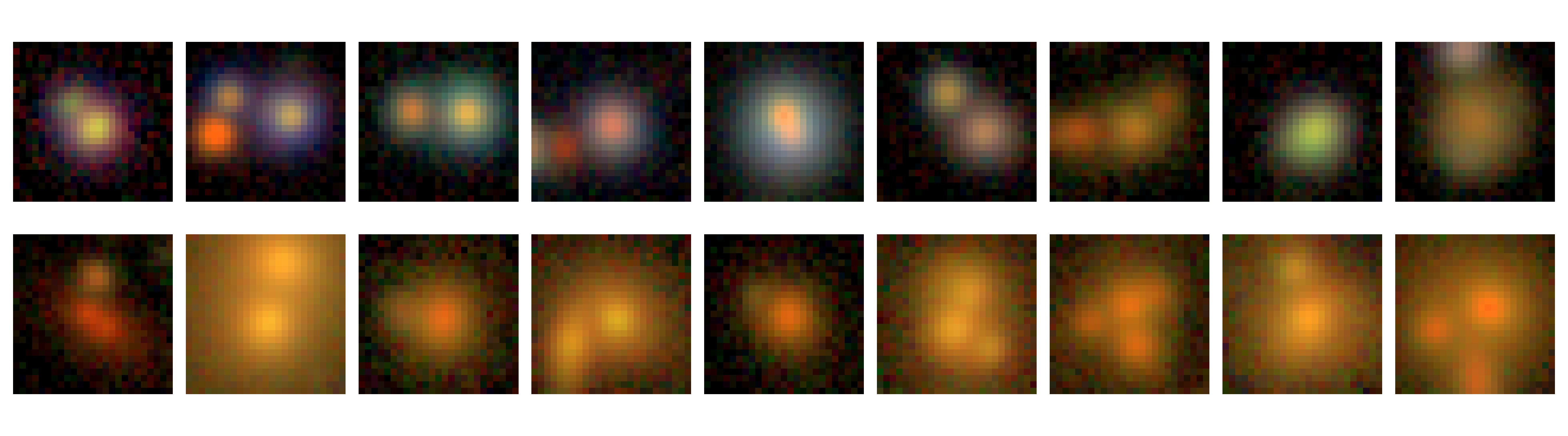}
\caption{Examples of false positives found by the Argus ensemble model. The false positives identified by Argus from a sample of lenses found by previous searches in the DES data are shown in the top row. The false positives identified by Argus from the simulated sample are shown in the bottom row.}
\label{fig:false_positives}
\end{figure*}

Figure~\ref{fig:false_positives} shows objects misclassified as lenses (false positives) by the Argus Ensemble model. Those in the top row were also selected by previous searches and were considered credible enough to warrant telescope time for investigation, so it is not catastrophic that they mislead Argus. Those in the bottom row were simulated false positives and indeed look like plausible lenses to a human classifier, so it does not seem catastrophic that they mislead Argus as well. At some level, a small number of false positive is inevitable due to the finite amount of information in the imaging data used for the search. Those false positives will need additional data to be resolved (spectroscopy and/or higher resolution images). We discuss the practical implications of this and other limitations to be kept in mind when applying this method to search for quads in the next section. 

\subsection{Practical Considerations}

\begin{figure}
\centering
\includegraphics[width=2.2in]{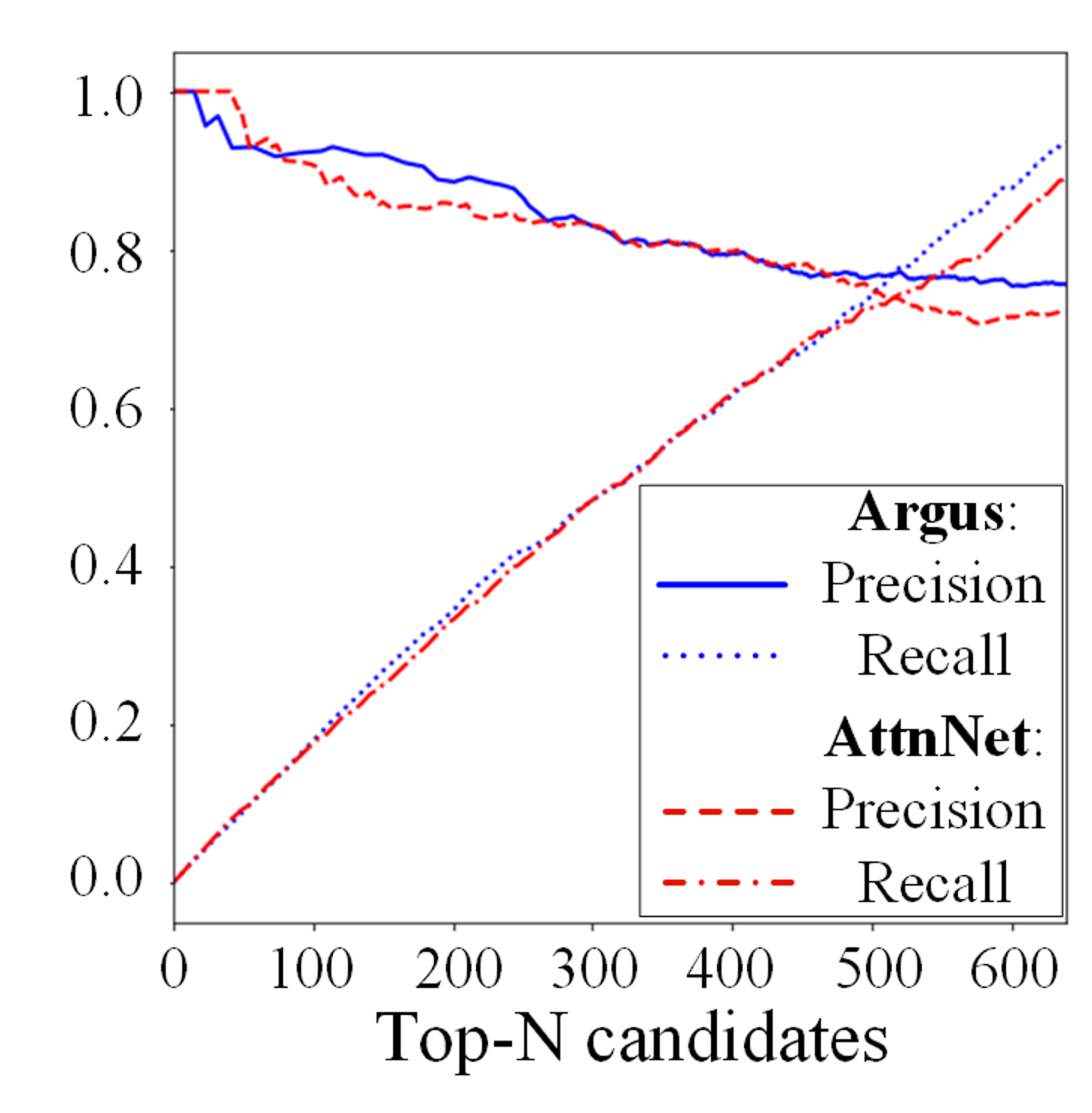}
\caption{Precision and recall at \textit{n} highest-ranked candidates in the \textit{mixed dataset}. The plots were generated by taking \textit{n} candidates with the highest scores estimated by the models as predicted \textit{positives} and taking the remaining objects as predicted \textit{negatives}.}
\label{fig:mixed_topn}
\end{figure}

Considering that telescope time is costly and limited, we wish to estimate how effectively lists of candidates produced by Argus or AttnNet can prioritize search time. The two key aspects of performance in this context are \textit{precision}---the proportion of candidates examined that prove to be lenses---and \textit{recall}, the fraction of true lenses that are identified in a given catalog. Figure~\ref{fig:mixed_topn} shows the estimated precision and recall as we adjust the number of selected candidates ranked by the models. 
Both Argus and AttnNet look promising in the sense that over a wide range of the number of candidates that might feasibly be considered, the precision rate remains quite high, well over 80\% until the number of candidates begins to approach the actual number of lenses in this sample (400). Naturally, the recall can at best grow linearly with a slope of 1, and the observed recall is not far below this. If these results generalize to images outside of those used in this paper, this plot would suggest that a search using our methods would be very efficient, yielding a complete sample of quadruply imaged quasars with a relatively modest fraction of contaminants. A follow-up paper will put our methods to further test by applying them to new DES datasets.


%
%

\section{Summary}
\label{sec:summary}

In this work, we proposed and trained novel deep learning-based methods for detecting quadruply lensed quasars. We investigated several ANN architectures and compared them with previously proposed LensFlow and CNNS models. Overall, our newly developed tools and in particular the ensemble model improve on the previously proposed algorithms \citep{AgneKelly15,WAT17,Hezaveh_fast_2017,Petrillo_KDS_2017,Petrillo_2018_KDS,CMU_deeplens_2017,Pourrahmani_lensflow_2018,Schaefer_2018,Avestruz_automated_2019,Madireddy_modular_2019,Cheng_unsupervised_2020,Jacobs:2019,Jacobs_candidatesDES_2019} in terms of precision and recall, although prior work was not tailored to the detection of quadruply imaged quasars, and therefore the comparison is not direct.

A major strength of our method is its ability to process $10^8$ objects in a single day on an ordinary hardware setup with a single GPU. This makes it suitable for discovering lenses across wide areas from surveys such as DES with minimal pre-screening of candidates, reducing the risk of discarding lenses. In a follow-up paper we will carry out such a search.
 
Two particularly novel and powerful elements of our approach are the use of polar convolution, which is well tuned to detecting circularly-arranged features at different scales, and attention masking. 
In the future, yet other ideas for feature extraction approaches, likely motivated by expert knowledge, may further improve performance. 

Finally, the challenges that remain are not only algorithmic, but in the acquisition of training data, real or simulated. That said, one area in which further progress could be made is in taking account of additional data, such as bands beyond griz, or using time series rather than single epoch images. 

\section*{Data availability statement}

All data and code required to reproduce these results and apply the trained model will be made publicly available upon publication at \url{https://des.ncsa.illinois.edu/releases/other/paper-data}. 

\section*{Acknowledgments}

TT acknowledges support from the National Science Foundation through grant NSF-AST-1906976, from NASA through grants HST-GO-15320 and HST-GO-15652, and from the Packard Foundation through a Packard Research Fellowship.

Funding for the DES Projects has been provided by the U.S. Department of Energy, the U.S. National Science Foundation, the Ministry of Science and Education of Spain, 
the Science and Technology Facilities Council of the United Kingdom, the Higher Education Funding Council for England, the National Center for Supercomputing 
Applications at the University of Illinois at Urbana-Champaign, the Kavli Institute of Cosmological Physics at the University of Chicago, 
the Center for Cosmology and Astro-Particle Physics at the Ohio State University,
the Mitchell Institute for Fundamental Physics and Astronomy at Texas A\&M University, Financiadora de Estudos e Projetos, 
Funda{\c c}{\~a}o Carlos Chagas Filho de Amparo {\`a} Pesquisa do Estado do Rio de Janeiro, Conselho Nacional de Desenvolvimento Cient{\'i}fico e Tecnol{\'o}gico and 
the Minist{\'e}rio da Ci{\^e}ncia, Tecnologia e Inova{\c c}{\~a}o, the Deutsche Forschungsgemeinschaft and the Collaborating Institutions in the Dark Energy Survey. 

The Collaborating Institutions are Argonne National Laboratory, the University of California at Santa Cruz, the University of Cambridge, Centro de Investigaciones Energ{\'e}ticas, 
Medioambientales y Tecnol{\'o}gicas-Madrid, the University of Chicago, University College London, the DES-Brazil Consortium, the University of Edinburgh, 
the Eidgen{\"o}ssische Technische Hochschule (ETH) Z{\"u}rich, 
Fermi National Accelerator Laboratory, the University of Illinois at Urbana-Champaign, the Institut de Ci{\`e}ncies de l'Espai (IEEC/CSIC), 
the Institut de F{\'i}sica d'Altes Energies, Lawrence Berkeley National Laboratory, the Ludwig-Maximilians Universit{\"a}t M{\"u}nchen and the associated Excellence Cluster Universe, 
the University of Michigan, NSF's NOIRLab, the University of Nottingham, The Ohio State University, the University of Pennsylvania, the University of Portsmouth, 
SLAC National Accelerator Laboratory, Stanford University, the University of Sussex, Texas A\&M University, and the OzDES Membership Consortium.

Based in part on observations at Cerro Tololo Inter-American Observatory at NSF's NOIRLab (NOIRLab Prop. ID 2012B-0001; PI: J. Frieman), which is managed by the Association of Universities for Research in Astronomy (AURA) under a cooperative agreement with the National Science Foundation.

The DES data management system is supported by the National Science Foundation under Grant Numbers AST-1138766 and AST-1536171.
The DES participants from Spanish institutions are partially supported by MICINN under grants ESP2017-89838, PGC2018-094773, PGC2018-102021, SEV-2016-0588, SEV-2016-0597, and MDM-2015-0509, some of which include ERDF funds from the European Union. IFAE is partially funded by the CERCA program of the Generalitat de Catalunya.
Research leading to these results has received funding from the European Research
Council under the European Union's Seventh Framework Program (FP7/2007-2013) including ERC grant agreements 240672, 291329, and 306478.
We  acknowledge support from the Brazilian Instituto Nacional de Ci\^encia
e Tecnologia (INCT) do e-Universo (CNPq grant 465376/2014-2).

This manuscript has been authored by Fermi Research Alliance, LLC under Contract No. DE-AC02-07CH11359 with the U.S. Department of Energy, Office of Science, Office of High Energy Physics.


\bibliography{ref_prior_works,references_ablai,BibdeskLib,references}

\begin{thebibliography}{}
\makeatletter
\relax
\def\mn@urlcharsother{\let\do\@makeother \do\$\do\&\do\#\do\^\do\_\do\%\do\~}
\def\mn@doi{\begingroup\mn@urlcharsother \@ifnextchar [ {\mn@doi@}
  {\mn@doi@[]}}
\def\mn@doi@[#1]#2{\def\@tempa{#1}\ifx\@tempa\@empty \href
  {http://dx.doi.org/#2} {doi:#2}\else \href {http://dx.doi.org/#2} {#1}\fi
  \endgroup}
\def\mn@eprint#1#2{\mn@eprint@#1:#2::\@nil}
\def\mn@eprint@arXiv#1{\href {http://arxiv.org/abs/#1} {{\tt arXiv:#1}}}
\def\mn@eprint@dblp#1{\href {http://dblp.uni-trier.de/rec/bibtex/#1.xml}
  {dblp:#1}}
\def\mn@eprint@#1:#2:#3:#4\@nil{\def\@tempa {#1}\def\@tempb {#2}\def\@tempc
  {#3}\ifx \@tempc \@empty \let \@tempc \@tempb \let \@tempb \@tempa \fi \ifx
  \@tempb \@empty \def\@tempb {arXiv}\fi \@ifundefined
  {mn@eprint@\@tempb}{\@tempb:\@tempc}{\expandafter \expandafter \csname
  mn@eprint@\@tempb\endcsname \expandafter{\@tempc}}}

\bibitem[\protect\citeauthoryear{{Agnello}, {Kelly}, {Treu}  \&
  {Marshall}}{{Agnello} et~al.}{2015}]{AgneKelly15}
{Agnello} A.,  {Kelly} B.~C.,  {Treu} T.,   {Marshall} P.~J.,  2015, \mn@doi
  [\mnras] {10.1093/mnras/stv037}, \href
  {http://adsabs.harvard.edu/abs/2015MNRAS.448.1446A} {448, 1446}

\bibitem[\protect\citeauthoryear{{Annis} et~al.,}{{Annis}
  et~al.}{2014}]{Annis:2014}
{Annis} J.,  et~al., 2014, \mn@doi [\apj] {10.1088/0004-637X/794/2/120}, \href
  {https://ui.adsabs.harvard.edu/abs/2014ApJ...794..120A} {794, 120}

\bibitem[\protect\citeauthoryear{Avestruz, Li, Zhu, Lightman, Collett  \&
  Luo}{Avestruz et~al.}{2019}]{Avestruz_automated_2019}
Avestruz C.,  Li N.,  Zhu H.,  Lightman M.,  Collett T.~E.,   Luo W.,  2019,
  \mn@doi [The Astrophysical Journal] {10.3847/1538-4357/ab16d9}, 877, 58

\bibitem[\protect\citeauthoryear{Cheng, Li, Conselice, Aragon-Salamanca, Dye
  \& Metcalf}{Cheng et~al.}{2020}]{Cheng_unsupervised_2020}
Cheng T.-Y.,  Li N.,  Conselice C.~J.,  Aragon-Salamanca A.,  Dye S.,   Metcalf
  R.~B.,  2020, \mn@doi [Monthly Notices of the Royal Astronomical Society]
  {10.1093/mnras/staa1015}, 494, 3750

\bibitem[\protect\citeauthoryear{Clevert, Unterthiner  \& Hochreiter}{Clevert
  et~al.}{2016}]{Clevert_ELU_ICLR_2016}
Clevert D.-A.,  Unterthiner T.,   Hochreiter S.,  2016, in Bengio Y.,  LeCun
  Y.,  eds, 4th International Conference on Learning Representations, {ICLR}
  2016, San Juan, Puerto Rico, May 2-4, 2016, Conference Track Proceedings.
  \url {http://arxiv.org/abs/1511.07289}

\bibitem[\protect\citeauthoryear{Doersch}{Doersch}{2016}]{vae_tutorial}
Doersch C.,  2016, Tutorial on Variational Autoencoders (\mn@eprint {arXiv}
  {1606.05908})

\bibitem[\protect\citeauthoryear{Goodfellow, Bengio  \& Courville}{Goodfellow
  et~al.}{2016}]{Goodfellow_DL_BOOK_2016}
Goodfellow I.,  Bengio Y.,   Courville A.,  2016, Deep Learning.
MIT Press

\bibitem[\protect\citeauthoryear{He, Zhang, Ren  \& Sun}{He
  et~al.}{2015}]{He_PRELU_ICCV_2015}
He K.,  Zhang X.,  Ren S.,   Sun J.,  2015, in Proceedings of the 2015 IEEE
  International Conference on Computer Vision (ICCV). ICCV 15.
IEEE Computer Society, USA, p. 1026–1034, \mn@doi{10.1109/ICCV.2015.123},
  \url {https://doi.org/10.1109/ICCV.2015.123}

\bibitem[\protect\citeauthoryear{He, Zhang, Ren  \& Sun}{He
  et~al.}{2016}]{He_DL_ResNet_2016}
He K.,  Zhang X.,  Ren S.,   Sun J.,  2016, in The IEEE Conference on Computer
  Vision and Pattern Recognition (CVPR).

\bibitem[\protect\citeauthoryear{Hezaveh, Levasseur  \& Marshall}{Hezaveh
  et~al.}{2017}]{Hezaveh_fast_2017}
Hezaveh Y.~D.,  Levasseur L.~P.,   Marshall P.~J.,  2017, \mn@doi [Nature]
  {10.1038/nature23463}, 548, 555

\bibitem[\protect\citeauthoryear{{Huang}, {Liu}, {Van Der Maaten}  \&
  {Weinberger}}{{Huang} et~al.}{2017}]{Huang_DL_DNET_2017}
{Huang} G.,  {Liu} Z.,  {Van Der Maaten} L.,   {Weinberger} K.~Q.,  2017, in
  2017 IEEE Conference on Computer Vision and Pattern Recognition (CVPR). pp
  2261--2269, \mn@doi{10.1109/CVPR.2017.243}

\bibitem[\protect\citeauthoryear{{Jacobs} et~al.,}{{Jacobs}
  et~al.}{2019a}]{Jacobs:2019}
{Jacobs} C.,  et~al., 2019a, \mn@doi [\apjs] {10.3847/1538-4365/ab26b6}, \href
  {https://ui.adsabs.harvard.edu/abs/2019ApJS..243...17J} {243, 17}

\bibitem[\protect\citeauthoryear{Jacobs et~al.,}{Jacobs
  et~al.}{2019b}]{Jacobs_candidatesDES_2019}
Jacobs C.,  et~al., 2019b, \mn@doi [The Astrophysical Journal Supplement
  Series] {10.3847/1538-4365/ab26b6}, 243, 17

\bibitem[\protect\citeauthoryear{Kingma \& Ba}{Kingma \&
  Ba}{2015}]{Kingma_DL_Adam_2015}
Kingma D.~P.,  Ba J.,  2015, in Bengio Y.,  LeCun Y.,  eds, 3rd International
  Conference on Learning Representations, {ICLR} 2015, San Diego, CA, USA, May
  7-9, 2015, Conference Track Proceedings. \url
  {http://arxiv.org/abs/1412.6980}

\bibitem[\protect\citeauthoryear{Klambauer, Unterthiner, Mayr  \&
  Hochreiter}{Klambauer et~al.}{2017}]{Klambauer_SELU_NIPS2017}
Klambauer G.,  Unterthiner T.,  Mayr A.,   Hochreiter S.,  2017, in Proceedings
  of the 31st International Conference on Neural Information Processing
  Systems. NIPS 17.
Curran Associates Inc., Red Hook, NY, USA, p. 972–981

\bibitem[\protect\citeauthoryear{Lanusse, Ma, Li, Collett, Li, Ravanbakhsh,
  Mandelbaum  \& Poczos}{Lanusse et~al.}{2017}]{CMU_deeplens_2017}
Lanusse F.,  Ma Q.,  Li N.,  Collett T.~E.,  Li C.-L.,  Ravanbakhsh S.,
  Mandelbaum R.,   Poczos B.,  2017, \mn@doi [Monthly Notices of the Royal
  Astronomical Society] {10.1093/mnras/stx1665}, 473, 3895

\bibitem[\protect\citeauthoryear{LeCun, Bengio  \& Hinton}{LeCun
  et~al.}{2015}]{lecun_DL_deeplearning_Nature_2015}
LeCun Y.,  Bengio Y.,   Hinton G.,  2015, \mn@doi [Nature]
  {10.1038/nature14539}, 521, 436

\bibitem[\protect\citeauthoryear{{Lemon} et~al.,}{{Lemon}
  et~al.}{2020}]{Lemon:2020}
{Lemon} C.,  et~al., 2020, \mn@doi [\mnras] {10.1093/mnras/staa652}, \href
  {https://ui.adsabs.harvard.edu/abs/2020MNRAS.494.3491L} {494, 3491}

\bibitem[\protect\citeauthoryear{Li, Jamieson, DeSalvo, Rostamizadeh  \&
  Talwalkar}{Li et~al.}{2018}]{hyperband}
Li L.,  Jamieson K.,  DeSalvo G.,  Rostamizadeh A.,   Talwalkar A.,  2018,
  Journal of Machine Learning Research, 18, 1

\bibitem[\protect\citeauthoryear{Madireddy, Li, Ramachandra, Butler,
  Balaprakash, Habib  \& Heitmann}{Madireddy
  et~al.}{2019}]{Madireddy_modular_2019}
Madireddy S.,  Li N.,  Ramachandra N.,  Butler J.,  Balaprakash P.,  Habib S.,
   Heitmann K.,  2019, A Modular Deep Learning Pipeline for Galaxy-Scale Strong
  Gravitational Lens Detection and Modeling (\mn@eprint {arXiv} {1911.03867})

\bibitem[\protect\citeauthoryear{{More} et~al.,}{{More}
  et~al.}{2016}]{More:2016}
{More} A.,  et~al., 2016, \mn@doi [\mnras] {10.1093/mnras/stv1965}, \href
  {https://ui.adsabs.harvard.edu/abs/2016MNRAS.455.1191M} {455, 1191}

\bibitem[\protect\citeauthoryear{{Oguri} \& {Marshall}}{{Oguri} \&
  {Marshall}}{2010}]{O+M10}
{Oguri} M.,  {Marshall} P.~J.,  2010, \mn@doi [\mnras]
  {10.1111/j.1365-2966.2010.16639.x}, \href
  {http://cdsads.u-strasbg.fr/abs/2010MNRAS.405.2579O} {405, 2579}

\bibitem[\protect\citeauthoryear{Petrillo et~al.,}{Petrillo
  et~al.}{2017}]{Petrillo_KDS_2017}
Petrillo C.~E.,  et~al., 2017, \mn@doi [Monthly Notices of the Royal
  Astronomical Society] {10.1093/mnras/stx2052}, 472, 1129

\bibitem[\protect\citeauthoryear{Petrillo et~al.,}{Petrillo
  et~al.}{2018}]{Petrillo_2018_KDS}
Petrillo C.~E.,  et~al., 2018, \mn@doi [Monthly Notices of the Royal
  Astronomical Society] {10.1093/mnras/sty2683}, 482, 807

\bibitem[\protect\citeauthoryear{Pourrahmani, Nayyeri  \& Cooray}{Pourrahmani
  et~al.}{2018}]{Pourrahmani_lensflow_2018}
Pourrahmani M.,  Nayyeri H.,   Cooray A.,  2018, \mn@doi [The Astrophysical
  Journal] {10.3847/1538-4357/aaae6a}, 856, 68

\bibitem[\protect\citeauthoryear{Ramachandran, Zoph  \& Le}{Ramachandran
  et~al.}{2018}]{swish_Ramachandran_ICLR_2018}
Ramachandran P.,  Zoph B.,   Le Q.,  2018. \url
  {https://arxiv.org/pdf/1710.05941.pdf}

\bibitem[\protect\citeauthoryear{Ronneberger, Fischer  \& Brox}{Ronneberger
  et~al.}{2015}]{Ronneberger_DL_UNET_2015}
Ronneberger O.,  Fischer P.,   Brox T.,  2015, in Navab N.,  Hornegger J.,
  Wells W.~M.,   Frangi A.~F.,  eds, Medical Image Computing and
  Computer-Assisted Intervention -- MICCAI 2015. Springer International
  Publishing, Cham, pp 234--241

\bibitem[\protect\citeauthoryear{Rosenblatt}{Rosenblatt}{1958}]{rosenblatt_perceptron_1958}
Rosenblatt F.,  1958, \mn@doi [Psychological Review] {10.1037/h0042519}, 65,
  386

\bibitem[\protect\citeauthoryear{Rozo et~al.,}{Rozo
  et~al.}{2016}]{Rozo_etal_redMaGic_2016}
Rozo E.,  et~al., 2016, \mn@doi [Monthly Notices of the Royal Astronomical
  Society] {10.1093/mnras/stw1281}, 461, 1431

\bibitem[\protect\citeauthoryear{{Schaefer, C.}, {Geiger, M.}, {Kuntzer, T.}
  \& {Kneib, J.-P.}}{{Schaefer, C.} et~al.}{2018}]{Schaefer_2018}
{Schaefer, C.} {Geiger, M.} {Kuntzer, T.}  {Kneib, J.-P.} 2018, \mn@doi [A\&A]
  {10.1051/0004-6361/201731201}, 611, A2

\bibitem[\protect\citeauthoryear{{Sevilla-Noarbe} et~al.,}{{Sevilla-Noarbe}
  et~al.}{2021}]{Sevilla-Noarbe2021}
{Sevilla-Noarbe} I.,  et~al., 2021, \mn@doi [\apjs] {10.3847/1538-4365/abeb66},
  \href {https://ui.adsabs.harvard.edu/abs/2021ApJS..254...24S} {254, 24}

\bibitem[\protect\citeauthoryear{{Szegedy} et~al.,}{{Szegedy}
  et~al.}{2015}]{Szegedy_DL_GoogleNet_2015}
{Szegedy} C.,  et~al., 2015, in 2015 IEEE Conference on Computer Vision and
  Pattern Recognition (CVPR). pp~1--9, \mn@doi{10.1109/CVPR.2015.7298594}

\bibitem[\protect\citeauthoryear{Szegedy, Ioffe, Vanhoucke  \& Alemi}{Szegedy
  et~al.}{2017}]{szegedy_DL_GoogleNet_v4_2017}
Szegedy C.,  Ioffe S.,  Vanhoucke V.,   Alemi A.~A.,  2017, in Proceedings of
  the Thirty-First AAAI Conference on Artificial Intelligence. AAAI 17.
AAAI Press, pp 4278--4284, \mn@doi{10.5555/3298023.3298188}

\bibitem[\protect\citeauthoryear{Tie et~al.,}{Tie et~al.}{2017}]{Tie2017}
Tie S.~S.,  et~al., 2017, \mn@doi [The Astronomical Journal]
  {10.3847/1538-3881/aa5b8d}, 153, 107

\bibitem[\protect\citeauthoryear{{Treu}}{{Treu}}{2010}]{Treu:2010}
{Treu} T.,  2010, \mn@doi [\araa] {10.1146/annurev-astro-081309-130924}, \href
  {https://ui.adsabs.harvard.edu/abs/2010ARA&A..48...87T} {48, 87}

\bibitem[\protect\citeauthoryear{{Treu} et~al.,}{{Treu}
  et~al.}{2018}]{Treu:2018}
{Treu} T.,  et~al., 2018, \mn@doi [\mnras] {10.1093/mnras/sty2329}, \href
  {https://ui.adsabs.harvard.edu/abs/2018MNRAS.481.1041T} {481, 1041}

\bibitem[\protect\citeauthoryear{{Vernardos}}{{Vernardos}}{2019}]{Vernardos:2019}
{Vernardos} G.,  2019, \mn@doi [\mnras] {10.1093/mnras/sty3486}, \href
  {https://ui.adsabs.harvard.edu/abs/2019MNRAS.483.5583V} {483, 5583}

\bibitem[\protect\citeauthoryear{{Wang}}{{Wang}}{2016}]{wang_DL_perspective_review_2016}
{Wang} G.,  2016, \mn@doi [IEEE Access] {10.1109/ACCESS.2016.2624938}, 4, 8914

\bibitem[\protect\citeauthoryear{{Williams}, {Agnello}  \& {Treu}}{{Williams}
  et~al.}{2017a}]{WAT17}
{Williams} P.,  {Agnello} A.,   {Treu} T.,  2017a, \mn@doi [\mnras]
  {10.1093/mnras/stw3239}, \href
  {http://adsabs.harvard.edu/abs/2017MNRAS.466.3088W} {466, 3088}

\bibitem[\protect\citeauthoryear{{Williams}, {Agnello}  \& {Treu}}{{Williams}
  et~al.}{2017b}]{Williams2017}
{Williams} P.,  {Agnello} A.,   {Treu} T.,  2017b, \mn@doi [\mnras]
  {10.1093/mnras/stw3239}, \href
  {https://ui.adsabs.harvard.edu/abs/2017MNRAS.466.3088W} {466, 3088}

\bibitem[\protect\citeauthoryear{{de Vaucouleurs}}{{de
  Vaucouleurs}}{1948}]{deV48}
{de Vaucouleurs} G.,  1948, Annales d'Astrophysique, \href
  {http://adsabs.harvard.edu/abs/1948AnAp...11..247D} {11, 247}

\makeatother
\end{thebibliography}
\bibliographystyle{mnras}

\appendix
\section{Affiliations}
\label{sec:appendix}

$^{1}$ Department of Electrical and Computer Engineering, University of California, Los Angeles, CA 90095, USA\\
$^{2}$ Department of Electrical and Computer Engineering, Nazarbayev University, Nur-Sultan, Kazakhstan\\
$^{3}$ The Inter-University Centre for Astronomy and Astrophysics (IUCAA), Post Bag 4, Ganeshkhind, Pune 411007, India\\
$^{4}$ Kavli IPMU (WPI), UTIAS, The University of Tokyo, Kashiwa, Chiba 277-8583, Japan\\
$^{5}$ Department of Statistics, University of California, Los Angeles, CA 90095, USA\\
$^{6}$ Department of Political Science, University of California, Los Angeles, CA 90095, USA\\
$^{7}$ Department of Physics and Astronomy, PAB, 430 Portola Plaza, Box 951547, Los Angeles, CA 90095-1547, USA\\
$^{8}$ Graduate School of Education, Stanford University, 160, 450 Serra Mall, Stanford, CA 94305, USA\\
$^{9}$ Department of Astronomy \& Astrophysics, University of Chicago, Chicago, IL 60637, USA\\
$^{10}$ MIT Kavli Institute for Astrophysics and Space Research, 37-664G, 77 Massachusetts Avenue, Cambridge, MA 02139, USA\\
$^{11}$ Laboratoire d'Astrophysique, Ecole Polytechnique F\'ed\'erale de Lausanne (EPFL), Observatoire de Sauverny, CH-1290 Versoix, Switzerland\\
$^{12}$ Fermi National Accelerator Laboratory, P. O. Box 500, Batavia, IL 60510, USA\\
$^{13}$ Kavli Institute for Cosmological Physics, University of Chicago, Chicago, IL 60637, USA\\
$^{14}$ Laborat\'orio Interinstitucional de e-Astronomia - LIneA, Rua Gal. Jos\'e Cristino 77, Rio de Janeiro, RJ - 20921-400, Brazil\\
$^{15}$ Instituto de F\'{i}sica Te\'orica, Universidade Estadual Paulista, S\~ao Paulo, Brazil\\
$^{16}$ Department of Physics \& Astronomy, University College London, Gower Street, London, WC1E 6BT, UK\\
$^{17}$ Kavli Institute for Particle Astrophysics \& Cosmology, P. O. Box 2450, Stanford University, Stanford, CA 94305, USA\\
$^{18}$ SLAC National Accelerator Laboratory, Menlo Park, CA 94025, USA\\
$^{19}$ Center for Astrophysical Surveys, National Center for Supercomputing Applications, 1205 West Clark St., Urbana, IL 61801, USA\\
$^{20}$ Department of Astronomy, University of Illinois at Urbana-Champaign, 1002 W. Green Street, Urbana, IL 61801, USA\\
$^{21}$ Institut de F\'{\i}sica d'Altes Energies (IFAE), The Barcelona Institute of Science and Technology, Campus UAB, 08193 Bellaterra (Barcelona) Spain\\
$^{22}$ Center for Cosmology and Astro-Particle Physics, The Ohio State University, Columbus, OH 43210, USA\\
$^{23}$ Jodrell Bank Center for Astrophysics, School of Physics and Astronomy, University of Manchester, Oxford Road, Manchester, M13 9PL, UK\\
$^{24}$ University of Nottingham, School of Physics and Astronomy, Nottingham NG7 2RD, UK\\
$^{25}$ Astronomy Unit, Department of Physics, University of Trieste, via Tiepolo 11, I-34131 Trieste, Italy\\
$^{26}$ INAF-Osservatorio Astronomico di Trieste, via G. B. Tiepolo 11, I-34143 Trieste, Italy\\
$^{27}$ Institute for Fundamental Physics of the Universe, Via Beirut 2, 34014 Trieste, Italy\\
$^{28}$ Observat\'orio Nacional, Rua Gal. Jos\'e Cristino 77, Rio de Janeiro, RJ - 20921-400, Brazil\\
$^{29}$ Department of Physics, University of Michigan, Ann Arbor, MI 48109, USA\\
$^{30}$ Hamburger Sternwarte, Universit\"{a}t Hamburg, Gojenbergsweg 112, 21029 Hamburg, Germany\\
$^{31}$ Centro de Investigaciones Energ\'eticas, Medioambientales y Tecnol\'ogicas (CIEMAT), Madrid, Spain\\
$^{32}$ Department of Physics, IIT Hyderabad, Kandi, Telangana 502285, India\\
$^{33}$ Faculty of Physics, Ludwig-Maximilians-Universit\"at, Scheinerstr. 1, 81679 Munich, Germany\\
$^{34}$ Santa Cruz Institute for Particle Physics, Santa Cruz, CA 95064, USA\\
$^{35}$ Institute of Theoretical Astrophysics, University of Oslo. P.O. Box 1029 Blindern, NO-0315 Oslo, Norway\\
$^{36}$ Instituto de Fisica Teorica UAM/CSIC, Universidad Autonoma de Madrid, 28049 Madrid, Spain\\
$^{37}$ Department of Astronomy, University of Michigan, Ann Arbor, MI 48109, USA\\
$^{38}$ Universit\"ats-Sternwarte, Fakult\"at f\"ur Physik, Ludwig-Maximilians Universit\"at M\"unchen, Scheinerstr. 1, 81679 M\"unchen, Germany\\
$^{39}$ School of Mathematics and Physics, University of Queensland,  Brisbane, QLD 4072, Australia\\
$^{40}$ Department of Physics, The Ohio State University, Columbus, OH 43210, USA\\
$^{41}$ Center for Astrophysics $\vert$ Harvard \& Smithsonian, 60 Garden Street, Cambridge, MA 02138, USA\\
$^{42}$ Lawrence Berkeley National Laboratory, 1 Cyclotron Road, Berkeley, CA 94720, USA\\
$^{43}$ Australian Astronomical Optics, Macquarie University, North Ryde, NSW 2113, Australia\\
$^{44}$ Lowell Observatory, 1400 Mars Hill Rd, Flagstaff, AZ 86001, USA\\
$^{45}$ Departamento de F\'isica Matem\'atica, Instituto de F\'isica, Universidade de S\~ao Paulo, CP 66318, S\~ao Paulo, SP, 05314-970, Brazil\\
$^{46}$ Department of Physics and Astronomy, University of Pennsylvania, Philadelphia, PA 19104, USA\\
$^{47}$ Instituci\'o Catalana de Recerca i Estudis Avan\c{c}ats, E-08010 Barcelona, Spain\\
$^{48}$ Physics Department, 2320 Chamberlin Hall, University of Wisconsin-Madison, 1150 University Avenue Madison, WI  53706-1390\\
$^{49}$ Institute of Astronomy, University of Cambridge, Madingley Road, Cambridge CB3 0HA, UK\\
$^{50}$ Department of Astrophysical Sciences, Princeton University, Peyton Hall, Princeton, NJ 08544, USA\\
$^{51}$ Institut d'Estudis Espacials de Catalunya (IEEC), 08034 Barcelona, Spain\\
$^{52}$ Institute of Space Sciences (ICE, CSIC),  Campus UAB, Carrer de Can Magrans, s/n,  08193 Barcelona, Spain\\
$^{53}$ School of Physics and Astronomy, University of Southampton,  Southampton, SO17 1BJ, UK\\
$^{54}$ Computer Science and Mathematics Division, Oak Ridge National Laboratory, Oak Ridge, TN 37831\\
$^{55}$ Department of Physics, Stanford University, 382 Via Pueblo Mall, Stanford, CA 94305, USA\\
$^{56}$ Max Planck Institute for Extraterrestrial Physics, Giessenbachstrasse, 85748 Garching, Germany\\

\label{lastpage}

\end{document}